\documentclass[10pt, pre, amsmath, onecolumn, showpacs, superscriptaddress, nofootinbib]{revtex4-1}

\usepackage{amsmath, amsthm, amssymb, amsfonts}
\usepackage{mathtools}
\usepackage{hyperref}
\hypersetup{
    colorlinks=true,       
    linkcolor=red,          
    citecolor=blue,        
    filecolor=magenta,      
    urlcolor=cyan           
}
\usepackage{graphicx}
\usepackage{dsfont}
\usepackage{fullpage}
\usepackage{color}
\usepackage{soul} 
\usepackage[title]{appendix}
\usepackage{tikz}
\usetikzlibrary{patterns}
\usetikzlibrary{shapes.multipart}
\usetikzlibrary{arrows}

\theoremstyle{definition}

\definecolor{labelkey}{cmyk}{.4,.2,0,0}

\newcommand{\be}{\begin{equation}}
\newcommand{\ee}{\end{equation}}
\newcommand{\beq}{\begin{equation}}
\newcommand{\eeq}{\end{equation}}
\newcommand{\bea}{\begin{eqnarray}}
\newcommand{\eea}{\end{eqnarray}}
\newcommand{\nn}{\nonumber}

\begin{document}

\title{
Run-and-tumble particles with 1D Coulomb interaction: the active jellium model and the non-reciprocal self-gravitating gas}

\author{L\'eo \surname{Touzo}}
\affiliation{Laboratoire de Physique de l'Ecole Normale Sup\'erieure, CNRS, ENS and PSL Universit\'e, Sorbonne Universit\'e, Universit\'e Paris Cit\'e, 24 rue Lhomond, 75005 Paris, France}
\author{Pierre Le Doussal}
\affiliation{Laboratoire de Physique de l'Ecole Normale Sup\'erieure, CNRS, ENS and PSL Universit\'e, Sorbonne Universit\'e, Universit\'e Paris Cit\'e, 24 rue Lhomond, 75005 Paris, France}

\date{\today}

\begin{abstract}

Recently we studied $N$ run-and-tumble particles in one dimension -- which switch with rate $\gamma$ between driving velocities $\pm v_0$ -- interacting via the long range 1D Coulomb potential (also called rank interaction), both in the attractive and in the repulsive case, with and without a confining potential. 
We extend this study in two directions. 
First we consider the same system, but inside a harmonic confining potential, which we call "active jellium". We obtain a parametric representation of the particle density in the stationary state 
at large $N$, which we analyze in detail. Contrary to the linear potential, there is always a steady-state where the density has a bounded support. 
However, we find that the model still exhibits transitions between phases with different behaviors of the density at the edges, ranging from a continuous decay to a jump, or even a shock (i.e. a cluster of particles, which manifests as a delta peak in the density). Notably, the
interactions forbid a divergent density at the edges, which may occur in the non-interacting case.
In the second part, we consider a non-reciprocal version of the rank interaction: the $+$ particles (of velocity $+v_0$) are attracted towards the $-$ particles (of velocity $-v_0$) with a constant force $b/N$, while the $-$ particles are repelled by the $+$ particles with a force of same amplitude.
In order for a stationary state to exist we add a linear confining potential. We derive an explicit expression for the stationary density
at large $N$, which exhibits an explicit breaking of the mirror symmetry with respect to $x=0$. This again shows the existence of several phases, 
which differ by the presence or absence of a shock at $x=0$, with one phase even exhibiting a vanishing density on the whole region $x>0$.
Our analytical results are complemented by numerical simulations for finite $N$.

\end{abstract} 

\maketitle

\tableofcontents

\section{Introduction}

Active particles are systems which transform external energy into directed motion, leading to interesting out-of-equilibrium dynamics.
The simplest model of an active, self propelled particle is the run-and-tumble particle (RTP).
It is driven by telegraphic noise \cite{HJ95,W02,ML17,kac74,Orshinger90}, and mimics the 
motion of E. Coli bacteria \cite{Berg2004,TailleurCates}. It also arises in quasi-1D channels
with staggered flows \cite{DoronCohen, ShapiraCohen}.
Even a single RTP, when submitted to an external trapping potential, reaches a non-trivial and
non-Boltzmann stationary state, where activity remains relevant at large times \cite{HJ95,Solon15, TDV16, DKM19, DD19,3statesBasu,LMS2020}. Due to the simplicity of the RTP models, analytical solutions are sometimes
possible. There is thus much to gain by studying them, as a starting point to elaborate more 
precise and detailed theories for
the many complex phenomena exhibited by active matter, see e.g. \cite{Berg2004,TailleurCates,soft,BechingerRev,Ramaswamy2017,Marchetti2018,Cates2012}. 
To this aim it is important to investigate what happens to these stationary states 
in the presence of interactions between the RTP's. 

Interacting active particles exhibit remarkable collective effects, such as motility-induced phase separation,  
clustering and jamming even for repulsive interactions and in the absence of alignment \cite{TailleurCates,soft,Ramaswamy2017,FM2012,Buttinoni2013,FHM2014,CT2015,slowman,slowman2,Active_OU,BG2021,CMPT2010,SG2014}. To describe the effects of interactions beyond numerical simulations, hydrodynamic approaches and perturbative exact results have been developed \cite{TailleurCates,Active_OU,KH2018,Agranov2021,Agranov2022}. However, there are at present very few exact results, even in one dimension, beyond the case of two interacting RTP's on the line \cite{slowman,slowman2,us_bound_state,Maes_bound_state,nonexistence,MBE2019,KunduGap2020,LMS2019}, or for 
harmonic chains \cite{SinghChain2020,PutBerxVanderzande2019, HarmonicChainRevABP,HarmonicChainRTPDhar}. Recently, exact solutions were also obtained for some specific many-particle models on a 1D lattice with contact interactions~\cite{Metson2022,MetsonLong,Dandekar2020,Thom2011}. In the continuum, an active version 
of the Dyson Brownian motion was introduced, where RTP's interact in 1D via a long range, logarithmic
potential, for which some analytical predictions were possible \cite{TouzoDBM2023}. 

Another type of long range interaction amenable to some exact results is the Coulomb potential in 1D, which is linear in the distance.
It is also called the rank interaction since the Coulomb force (either attractive or repulsive) acting on each particle is proportional to its rank, i.e., the number of particles in front of it, minus the number of particles behind. It has been much studied in the case of Brownian particles, in mathematics \cite{OConnell,Pitman}, in finance \cite{Banner} and in physics.
In physics, most studies have addressed the equilibrium steady state, known as the self-gravitating 1D gas
in the attractive case \cite{Rybicki,Sire,Kumar2017}, and as the Jellium model in the repulsive case in presence of a harmonic confining potential 
\cite{Lenard,Prager,Baxter,Dean1,Tellez,Lewin,SatyaJellium1,SatyaJellium2,SatyaJellium3,Flack22,Chafai_edge}. 
The non-equilibrium dynamics of this model was studied more recently in \cite{PLDRankedDiffusion,FlackRD} 
using connections with the Lieb-Liniger delta Bose gas model and the Burgers equation. 

Recently we have considered an active generalization of this model, 
consisting in $N$ RTP's in 1D mutually interacting via the linear Coulomb potential \cite{activeRDshort}.
It was solved previously for $N=2$ leading to a two particle stationary bound state in the attractive case \cite{us_bound_state}. In the limit $N\to+\infty$, we showed that the evolution of the density fields is described
by two coupled Burger's type equations. In the attractive case, and in the absence of external potential, these equations admit an exact stationary solution, which
describes a $N$-particle bound state. This bound state was found to exhibit transitions between (i) a phase 
where the density is smooth with infinite support, (ii) a phase where the density has finite support and exhibits "shocks", i.e. clusters of particles, 
at the edges, and (iii) a fully clustered phase.
The formation of these clusters is possible because of the non-analytic behavior of the Coulomb potential at the origin.
Adding a linear external potential, both in the attractive and the repulsive case, makes the phase diagram even richer: additional partially expanding phases appear, with or without shocks.

The aim of the present paper is two-fold. In the first part, we want to extend the result of this previous work to
a more natural confining potential, namely the harmonic well, leading to an "active jellium model" (which we will
consider both for repulsive and attractive interactions). 
Given
the form of the interaction, the linear potential was easier to treat analytically. The harmonic
potential is more difficult to analyze. We are still able to obtain a parametric representation
of the stationary density, but which is less explicit than for the linear potential. Nevertheless
we are able to extract a lot of information about the stationary state, such
as the phase diagram, the size of the support, the behavior of the densities near the
edges, as well as the presence or not of shocks, i.e. of clusters of particles. 
These predictions are tested
through numerical simulations. 

The second part addresses the case of non-reciprocal interactions between right moving ($+$ particles) and left moving particles ($-$ particles).
As an example we consider the case where the $+$ particles are attracted to the $-$ particles, while the $-$ particles are repelled by the $+$ particles (and two particles with the same sign do not interact together), which we call the "non-reciprocal active self-gravitating gas". Indeed there has been much recent interest in collective effects in systems with non-reciprocal interactions with remarkable phenomena such as pattern formation, synchronization and flocking, see e.g. \cite{VitelliNature2021}. Non-reciprocal interactions are relevant in the context of animal behavior \cite{nonreciprocal_active_book}, and have been studied experimentally in bacteria \cite{Kocabas2024}, but can also be observed at smaller scales \cite{Mandal2024}. Recently, models of active particles with various form of non-reciprocal interactions (from simple forces to alignment interactions and quorum sensing) have been studied either numerically or through the analysis of general field theories or hydrodynamic equations obtained by coarse-graining of a microscopic model \cite{You2020,nonreciprocalActive,KK2022,Knezevic2022,Dinelli2023,Duan2023,Duan2024,Du2024,Newman2008,Peruani2016,Peruani2017,Durve2018,Negi2022,Qi2022,Stengele2022,Saavedra2024}. These systems often exhibit very rich phase diagrams with phase separations, pattern formation and oscillating phases. These models often involve two or more distinct "species" of active particles between which the non-reciprocal interactions take place \cite{You2020,nonreciprocalActive,KK2022,Knezevic2022,Dinelli2023,Duan2023,Duan2024,Du2024}. There are however cases where the non-reciprocal interaction instead depends on the current "state" of the particle (such as its orientation, $+$ or $-$ in the present model), in particular in models with vision cones were particles only interact with the particles inside a certain angle with respect to their orientation \cite{Newman2008,Peruani2016,Peruani2017,Durve2018,Negi2022,Qi2022,Stengele2022,Saavedra2024}. The effect of such interactions on flocking in particular has attracted a lot of attention. The model considered here is somewhat closer to the second category, since the non-reciprocal interaction depends on the state of the particle, which changes with time. This choice is motivated by the objective to keep the model as analytically tractable as possible. It allows us to obtain an exact analytical solution for the stationary state and to obtain the phase diagram.
We find that here the non-reciprocity leads to a breaking of the spatial parity symmetry. 
Finally, as discussed above, a peculiar feature of the Coulomb interaction in one dimension is the appearance of shocks,
both in the passive and active case. 
It it thus an interesting question to investigate how some amount of non-reciprocity in the $+/-$ interaction
will affect the structure of the shocks. We find that in presence of a confining linear potential the non-reciprocity 
produces phases with shocks which where absent in the reciprocal case. 

We have also considered a different model with non-reciprocal interactions, more directly inspired from the notion of vision cone, where the particles only receive a force (still independent of the distance) from the particles "in front" of them (i.e. on the right for $+$ particles and on the left for $-$ particles). It turns out that this model can be mapped to the reciprocal active rank diffusion model studied in \cite{activeRDshort} and in the first half of this paper, so that the non-reciprocality does not play any particular role in this case. The details are given in Appendix~\ref{app:vision_cone}.

\section{Main results} \label{sec:mainresults}

\subsection{Rank interaction with a harmonic external potential}

We consider the active rank diffusion model introduced in \cite{activeRDshort}
\be \label{langevin1}
\frac{dx_i}{dt} = \frac{\kappa}{N} \sum_{j=1}^N {\rm sgn}(x_i-x_j) - V'(x_i) + v_0 \sigma_i(t) + \sqrt{2 T} \xi_i(t) \;.
\ee 
This model was studied in \cite{activeRDshort} for $V(x)=0$ and for $V(x)=a |x|$ with $a>0$, in the absence of thermal noise ($T=0$), where a complete analytical solution was obtained for the stationary state when it exists, or for the large time scaling form otherwise, 
both in the repulsive case $\kappa>0$ and in the attractive case $\kappa=-\bar \kappa <0$. For sufficiently strong 
attraction all the particles are grouped in a single cluster. For weaker attraction they either form a phase with a smooth density, which extends to infinity, or a phase where the density has a finite support and with "shocks" at the edges, i.e. clusters of same sign particles forming delta peaks in the density. For sufficiently repulsive interactions, $\kappa>a$, a finite fraction of the particles escape to infinity.

Here we will consider the case of a harmonic external potential 
\be 
V(x)=\frac{\mu}{2}x^2  \label{harmonic} 
\ee 
(again at $T=0$). The model \eqref{langevin1} then has two dimensionless parameters, namely
\begin{equation} \label{2params}
\frac{\mu}{\gamma} \quad {\rm and} \quad \frac{\kappa}{v_0} \;.
\end{equation}
Although we cannot obtain a fully explicit solution in all cases, some analytical progress can still be made.
Let us recall that for $v_0=0$, i.e. in the passive case, and $\kappa>0$, the equilibrium density at $T=0$ for 
sufficiently confining convex potentials is given by 
$\rho_{eq}(x)= V''(x)/(2 \kappa)$ (within a finite support), while for $T>0$ and for a 
harmonic potential it is given by Eqs. (27)-(28) in \cite{PLDRankedDiffusion}. 

Here we
study the time dependent density fields $\rho_\sigma(x,t)$ with $\sigma= \pm 1$
and their
even and odd components $\rho_s$ and $\rho_d$, defined as
\bea
&& \rho_\sigma(x,t) = \frac{1}{N} \sum_i \delta(x_i(t)-x) \delta_{\sigma_i(t),\sigma} \;, \\
&& \rho_{s/d}(x,t)=\rho_+(x,t) \pm \rho_-(x,t) \,. 
\eea

Since the potential is confining and the amplitude of the noise and interaction strength are finite, these densities will always have a finite support $[-x_e,x_e]$. 

Let us first recall the case without interactions $\kappa=0$. In that case it is known that the stationary densities 
take the form \cite{DKM19}
\be 
\rho_s(x) 
= \frac{2}{4^{\gamma/\mu} B(\frac{\gamma}{\mu},\frac{\gamma}{\mu})} \frac{\mu}{v_0} \left(1 - \left(\frac{\mu x}{v_0}\right)^2\right)^{\frac{\gamma}{\mu}-1} \quad , \quad \rho_d(x) = \frac{\mu x}{v_0} \rho_s(x) \;,
\ee 
where $B(\alpha, \beta)= \Gamma(\alpha) \Gamma(\beta)/\Gamma(\alpha+\beta)$ is the beta function.
This corresponds to
\be
\rho_\pm(x) = \frac{1}{4^{\gamma/\mu} B(\frac{\gamma}{\mu},\frac{\gamma}{\mu})} \frac{\mu}{v_0} \left(1 \pm \frac{\mu x}{v_0}\right)^{\frac{\gamma}{\mu}} \left(1 \mp \frac{\mu x}{v_0}\right)^{\frac{\gamma}{\mu}-1} \;.
\ee
Hence the densities are singular at the edges: the total density diverges for $\gamma<\mu$, 
vanishes for $\gamma>\mu$, and is uniform on the interval $[-x_e,x_e]$ for the marginal case $\gamma=\mu$, with $x_e=v_0/\mu$ in all cases. 
Note also that the density $\rho_+$ always vanishes at the left edge, and $\rho_-$ at the right edge. 

In this paper we study the $N\to+\infty$ limit of the model \eqref{langevin1} with the harmonic potential 
\eqref{harmonic}. The passive case $v_0=0$, $T>0$ was studied in \cite{PLDRankedDiffusion}, where the stationary density was found to take a scaling form
\be
\rho_{\rm eq} (x) = \frac{\mu}{2\kappa} \hat \rho_g \left( x\sqrt{\frac{\mu}{T}} \right) \quad , \quad g = \frac{\kappa}{\sqrt{\mu T}} \;,
\ee
where $\hat \rho_g(y)$ is a smooth function whose support is the whole real axis. As $g$ varies, it interpolates between a Gaussian in the weakly interacting limit $g\to 0$ and a square density in the small temperature limit $g\to+\infty$.

\begin{figure}[t]
    \centering
    \includegraphics[width=0.49\linewidth,trim={1.8cm 3cm 0.95cm 4.4cm},clip]{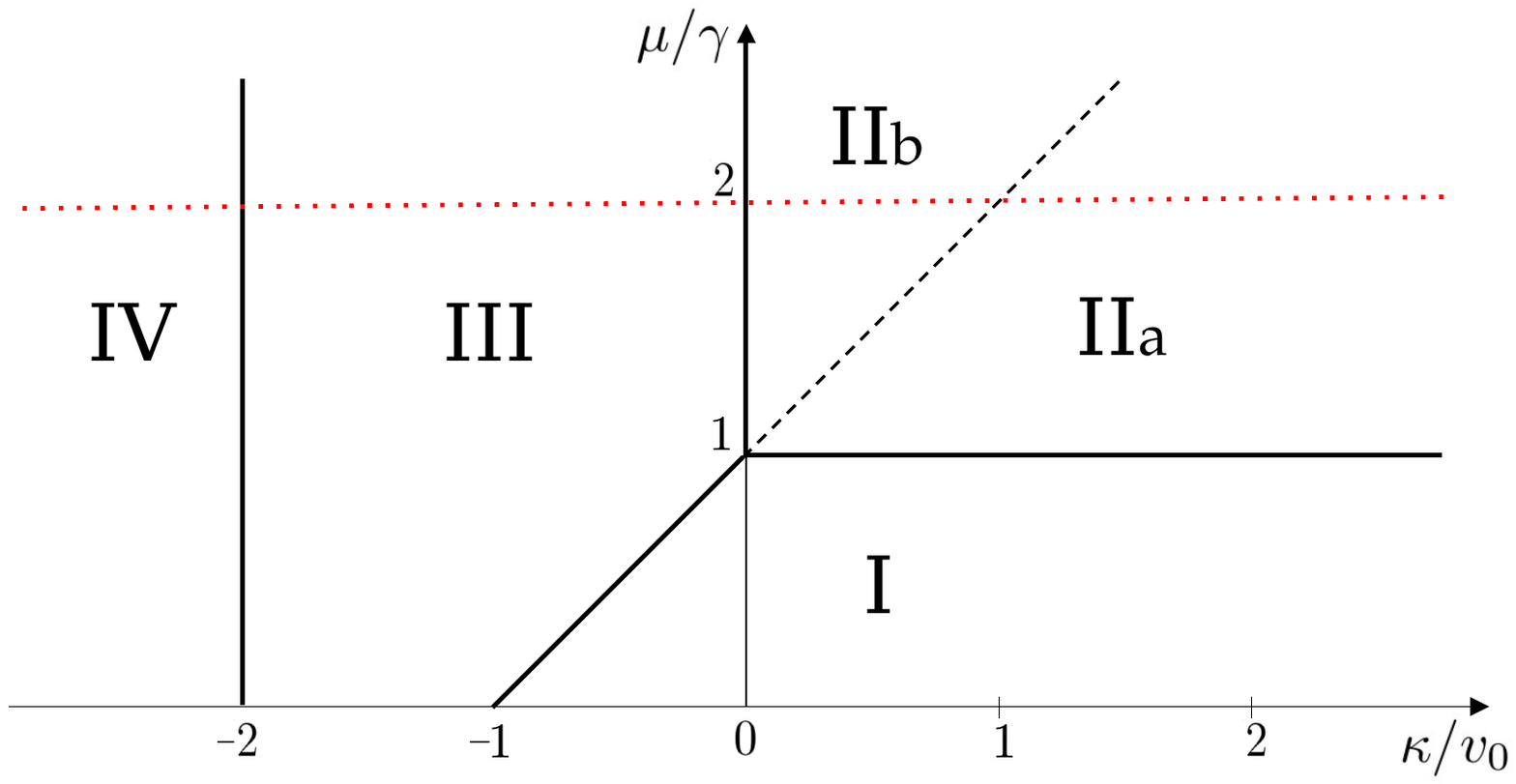}
    \includegraphics[width=0.49\linewidth,trim={0.95cm 4.55cm 1cm 3.6cm},clip]{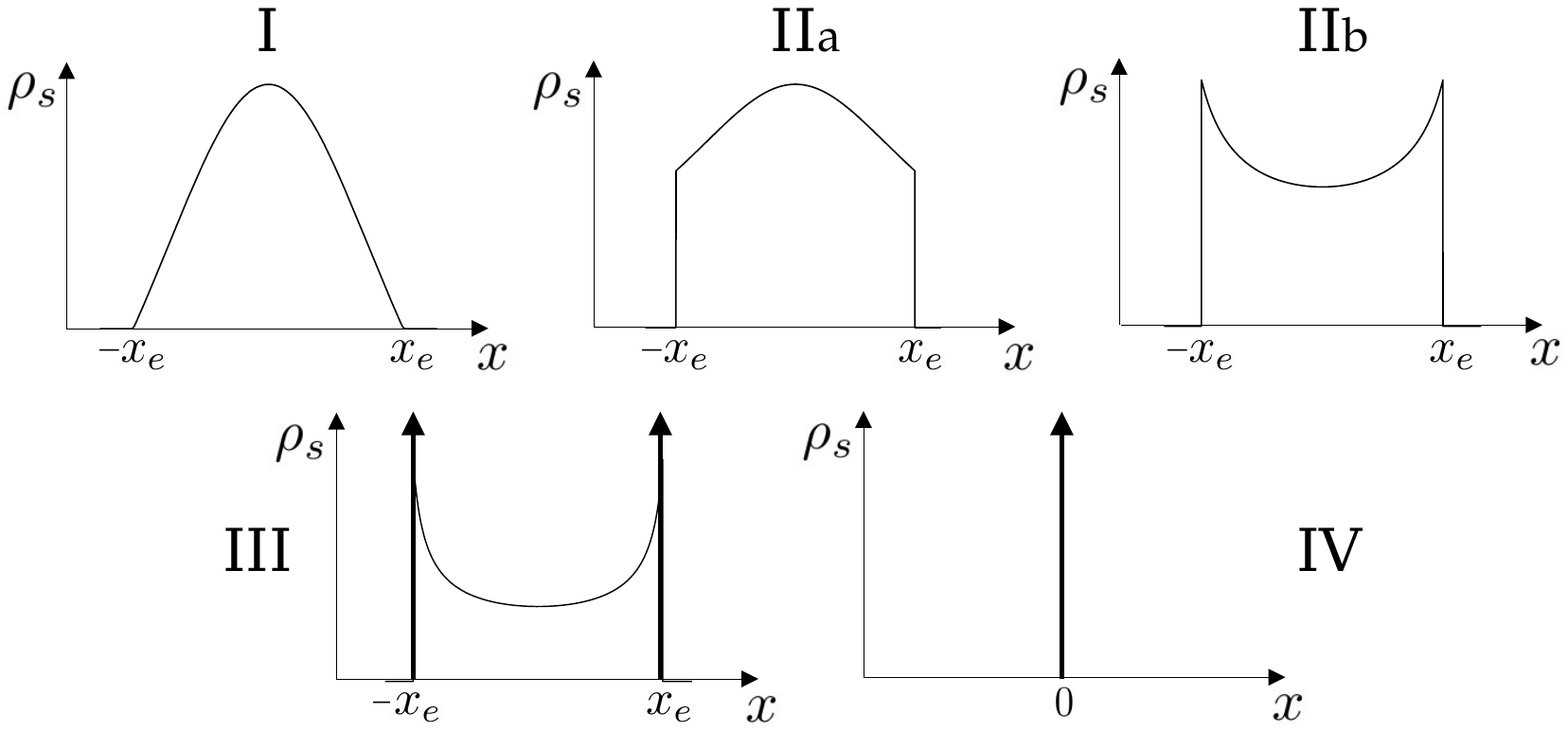}
    \caption{Left panel: phase diagram of the active jellium model, (i.e. the active rank diffusion in a harmonic potential $V(x)=\frac{\mu}{2} x^2$) in the plane $(\frac{\kappa}{v_0}, \frac{\mu}{\gamma})$, with repulsive interactions for $\kappa>0$ and attractive for $\kappa<0$. Right panel : 
    schematic representation of the total density $\rho_s(x)$ for each phase. The up arrows represent delta functions
    in the density, i.e. shocks/clusters of particles. In phases I, IIa and IIb there are no shocks. In phase I (which extends
    on either side of $\kappa=0$) the density
    vanishes at the edge (with an exponent $\frac{\gamma}{\mu}-1$ identical to the non-interacting case $\kappa=0$). In phases
    IIa and IIb the density has a finite jump at the edge (except for $\kappa=0$ where it diverges). The dotted line between phases IIa and IIb represents only a crossover where the density changes from concave to convex (not a true phase transition). In phase III there are
    shocks at the two edges, with a cluster of $+$ only particles at the right edge, and $-$ only at the left edge. In phase IV
    all the particles belong to a single cluster. In the text more explicit expressions are obtained on the special line $\frac{\mu}{\gamma}=2$, represented as a dotted red line on the phase diagram.}
    \label{phase_diagram}
\end{figure}

Here we focus on the active case $T=0$, $v_0>0$.
Using the same method as in \cite{activeRDshort}, we obtain a pair of partial differential equations for the rank fields, defined as
\be \label{rankdef1}
r(x,t) = \int^x_{-\infty} dy \, \rho_s(y,t)\, -\, \frac{1}{2} \; , \; s(x,t) = \int_{-\infty}^x dy \, \rho_d(y,t) \; .
\ee
We focus on the stationary state where $r(x,t)\to r(x)$ and $s(x,t)\to s(x)$. We then analyze the stationary equations to extract relevant information on the stationary densities $\rho_s(x)=r'(x)$ and $\rho_d(x)=s'(x)$. Since the potential is even in $x$, we find stationary solutions with $r(x) \in [-1/2,1/2]$ an odd function of $x$, and $s(x)$ an even function of $x$.
Our results are summarized in Fig.~\ref{phase_diagram}, which shows the different regimes as a function of the two dimensionless parameters \eqref{2params}, as well as a sketch of the total density $\rho_s(x)$ in each regime. 

\begin{itemize}
  
   \item In phase I, which extends both for repulsive and attractive interactions, 
   corresponding to $\frac{\mu}{\gamma}<\min (1,1+\frac{\kappa}{v_0})$ the total density $\rho_s(x)$ is smooth and vanishes at the edges with an exponent $\frac{\gamma}{\mu}-1$, similar to the non-interacting case $\kappa=0$. 
In fact, all the densities $\rho_\pm$ vanish with the same exponents as in the absence of interactions (although with different amplitudes). 

\item 
In phase II, which corresponds to the repulsive case with $\mu>\gamma$, 
the density instead converges to a finite value at the edges, leading to a jump of magnitude
\begin{equation}
\rho_s(x_e)=\frac{\mu-\gamma}{2\kappa} \;.
\end{equation}
Hence we find that the repulsive interactions suppress the divergence which is present for $\kappa=0$.
Furthermore, the density $\rho_+$ still vanishes at the left edge, and $\rho_-$ at the right edge, 
but they now vanish linearly as $|x \pm x_e|$, instead of superlinearly. 
The phase II is divided in two regimes IIa and IIb depending on the convexity of the density inside the support. 
In both phases I and II there are no shocks, and the support is given by $[-x_e,x_e]$ with
\begin{equation} \label{xe_intro}
    x_e=\frac{v_0+\kappa}{\mu} \;.
\end{equation}
On the boundary between the regimes I and IIa, i.e. for $\gamma=\mu$, the density
vanishes at the edge as 
\be
\rho_s(x) \simeq \frac{\mu}{2\kappa} \frac{1}{\big(\ln (x_e-x)\big)^2} \;.
\ee
This inverse logarithmic divergence is purely an effect of the interactions, as it occurs only for $\kappa >0$, while in the non-interacting
case the density is uniform for $x \in [-v_0/\mu,v_0/\mu]$ for $\gamma=\mu$.

\item
Phase III is restricted to the attractive case, for $\frac{\mu}{\gamma}>1+\frac{\kappa}{v_0}$ and $-2v_0<\kappa= - \bar \kappa <0$, and it is characterized by the presence of shocks at the edges, i.e. the density has delta peaks at $\pm x_e$, corresponding to a cluster of $+$ particles at $x_e$ and $-$ particles at $-x_e$. A similar phenomenon was observed both in the absence of a potential, and in the case of the linear potential. An important difference is that here one has $x_e>\frac{v_0-\bar\kappa}{\mu}$, i.e. the support is {\it extended} by the presence of shocks:
\be 
x_e = \frac{v_0-\bar \kappa}{\mu} + \frac{\bar \kappa p}{2 \mu}  
\ee 
where $p$ is the total fraction of particles in the two clusters. In the other two cases the support was infinite in the absence of shocks.

\item 
Finally, phase IV corresponds to the case $\kappa<-2v_0$, for which one trivially has $\rho_s(x)=\delta(x)$,
i.e. all particles belong to a single cluster, which is stable. 

\end{itemize}

\begin{figure}[t]
    \centering
    \includegraphics[width=0.32\linewidth]{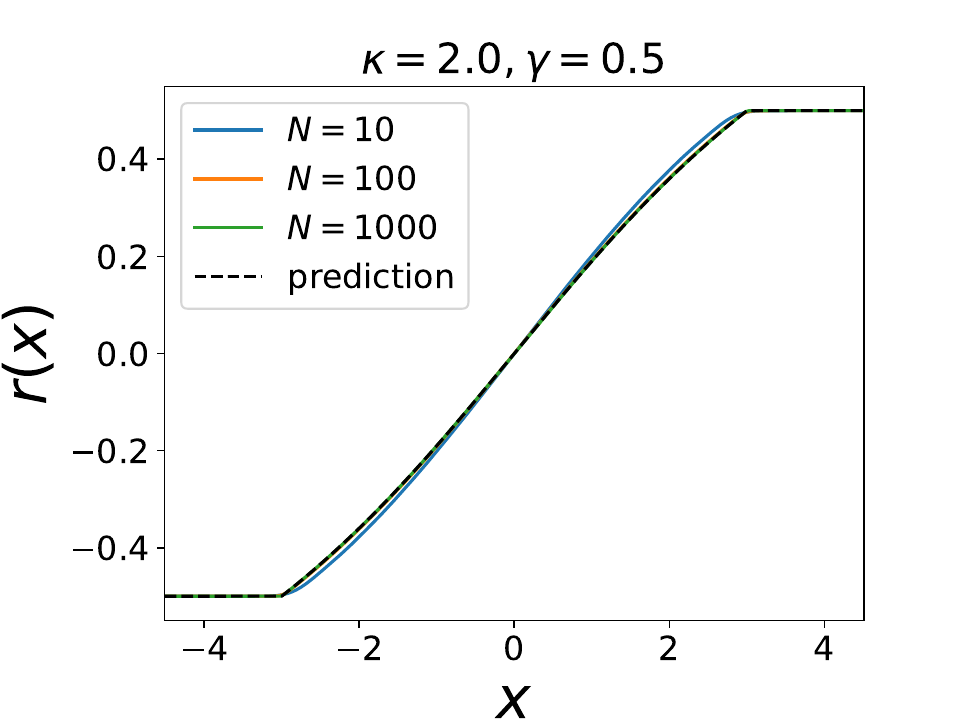}
    \includegraphics[width=0.32\linewidth]{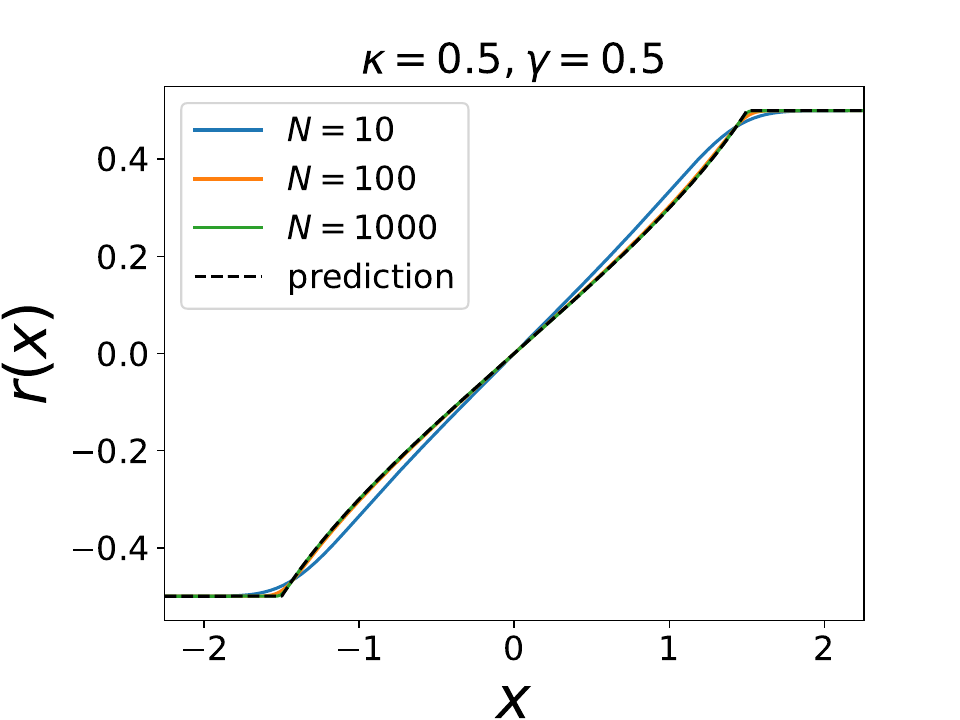}
    \includegraphics[width=0.32\linewidth]{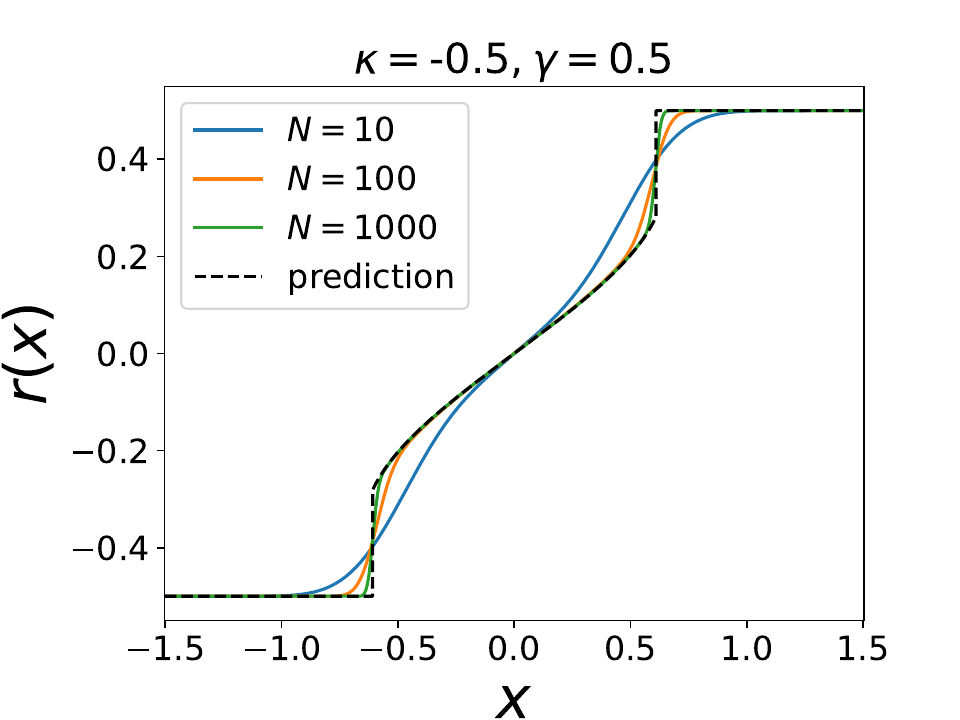}
    \includegraphics[width=0.32\linewidth]{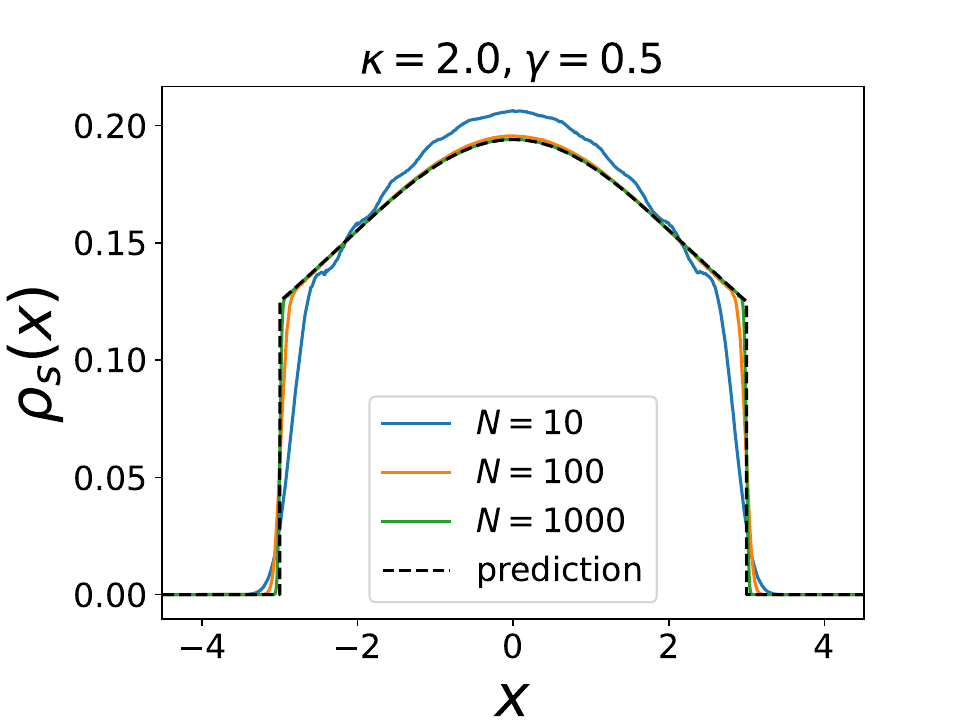}
    \includegraphics[width=0.32\linewidth]{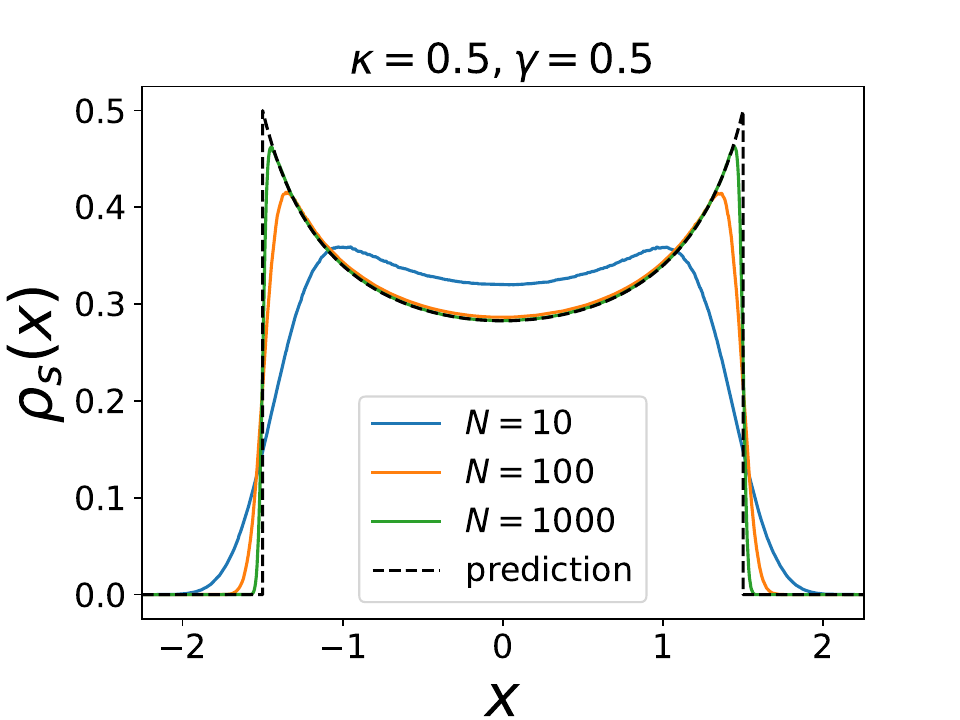}
    \includegraphics[width=0.32\linewidth]{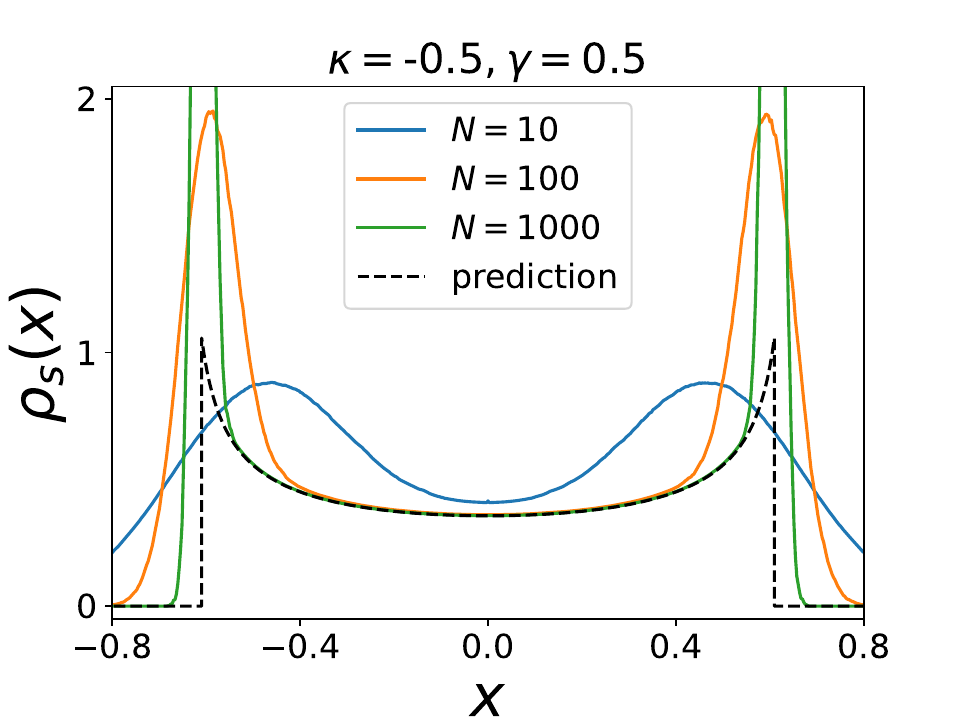}
    \caption{{\bf Top:} Comparison of the rank field $r(x)$ in the stationary state computed using numerical simulations with different values of $N$, with the analytical prediction for $a=1/2$ i.e. $\mu=2 \gamma$. In all cases $v_0=1$, $\mu=1$ and $\gamma=0.5$, and $\kappa$ varies to explore the 3 non-trivial regimes: $\kappa>v_0$, i.e. phase IIb (left), $0<\kappa<v_0$, i.e. phase IIa (center) and $-2v_0<\kappa<0$, i.e. phase III (right). In $r(x)$, the shocks appear as jumps for $N\to+\infty$ in the right panel. The dashed black lines correspond to the predictions of Eqs.~\eqref{r_repulsive2_intro}, \eqref{r_repulsive1_intro} and \eqref{r_repulsive1_intro}-\eqref{eq_edge_attractive_intro} respectively. {\bf Bottom:} Plot of the density $\rho_s(x)=r'(x)$ for the same parameters, obtained through numerical simulations. The dashed black lines correspond to the prediction \eqref{rhos_param} (the delta peaks are not shown on the right panel).}
    \label{fig_a1}
\end{figure}

We were also able to take the computations further in the special case $\mu/\gamma=2$ (the red line in 
Fig.~\ref{phase_diagram}).
For $\kappa>v_0$, i.e within phase IIa, the rank field $r(x)$ is obtained in parametric form
\be \label{r_repulsive2_intro}
\frac{\mu x}{v_0} \sinh(c/2)  = \tilde g_c(r) \quad , \quad \tilde g_c(r) :=\sinh (cr) + c \, r \cosh(\frac{c}{2}) \;,
\ee 
where $c\in[0,+\infty]$ is the solution of 
\be 
\frac{\tanh(c/2)}{c/2}= \frac{v_0}{\kappa} \;.
\ee 
Since $\tilde g_c(r)$ is an increasing function of $r$ for any value of $c$, the Eq. \eqref{r_repulsive2_intro} is invertible,
$r(x)$ is smooth and there is no shock. In that case the edge is given by the general formula \eqref{xe_intro}. 

For $\kappa<v_0$, we find instead
\be \label{r_repulsive1_intro}
\frac{\mu x}{v_0} \sin(c/2)  = g_c(r) \quad , \quad g_c(r) :=\sin (cr) + c\, r \cos(\frac{c}{2}) \quad , \quad \frac{\tan(c/2)}{c/2}= \frac{v_0}{\kappa} \;.
\ee 
In the repulsive case, $\kappa>0$, i.e. in phase IIb, one has $c\in[0,\pi]$ and $g_c(r)$ is again an increasing function on $[0,1/2]$, so that this relation can be inverted and there is no shock. The edge is again given by \eqref{xe_intro}.  In the attractive case, $\kappa<0$, i.e. in phase III, one has instead $c\in[\pi,2\pi]$ and $g_c(r)$ is no longer increasing on the whole interval $[0,1/2]$. In that case, we find
that there is a shock at the edge, i.e. a delta function in the density with weight $\frac{1}{2} - r(x_e^-)$.
This weight is determined from the equation
\be \label{eq_edge_attractive_intro}
h_c(r(x_e^-)) = h_c\big(\frac{1}{2}\big) \quad , \quad h_c(r)=\sin (cr) + \frac{cr}{2} \cos \big(\frac{c}{2}\big) \;.
\ee
The position of the edge $x_e$ is then obtained by inserting the value of $r(x_e^-)$ in \eqref{r_repulsive1_intro}.
More details on the derivations and the results are given below. 

Another interesting case is the limit $\gamma \ll \mu$. In that limit the system spends most of the time near "fixed points" of
the dynamics \cite{TouzoDBM2023}. We show that in the repulsive case $\kappa>0$, it results in the following bimodal shape for the total density 
\be
\rho_s(x) = \frac{\mu}{2\kappa} \quad {\rm for} \ \frac{v_0}{\mu}<|x|<\frac{v_0+\kappa}{\mu} \;,
\ee
i.e. the density exhibits a gap inside which it vanishes. This is explained by the fact that there is a separation between $+$ particles on the right and $-$ on the left. 

Finally in the diffusive limit $v_0 \to +\infty$, $\gamma \to + \infty$ with $T_a=\frac{v_0^2}{2 \gamma}$ fixed 
one can show (see \cite{activeRDshort}) from the equations derived below that at large $N$ the total density $\rho_s(x)$ converges to
the one of the passive case with $T=T_a$, given in \cite{PLDRankedDiffusion}. 

The above results, valid for $N \to +\infty$, have been tested by numerical simulations for large 
values of $N$. In Fig.~\ref{fig_a1}, the predictions for the rank field $r(x)$ and the total density $\rho_s(x)$ for $\mu=2\gamma$ in the 3 different regimes are compared with the results obtained by simulating the dynamics of equation \eqref{langevin1} and averaging over time, in the stationary state, for different values of $N$. At finite $N$ the density is always smooth (the jumps and the delta peaks at the edges have a finite width which decreases with $N$, due in the second case to fluctuations in the size and position of the clusters), but as $N$ increases the agreement with the predictions becomes very good. For more details on the method used for the numerical simulations, see the supplementary material of \cite{activeRDshort}.  

In Appendix~\ref{app:vision_cone}, we introduce a variant of the present model with non-reciprocal interactions inspired from the notion of vision cone. The interaction still takes the form of a 1D Coulomb potential, but now each particle only receives a force from the particles "in front" of it (i.e. on the right for $+$ particles and on the left for $-$ particles). It turns out that this model can be mapped to the reciprocal active rank diffusion model studied here in the case of a harmonic confining potential, and in \cite{activeRDshort} in the absence of confinement or with a linear potential. The full phase diagram for this model with non-reciprocal interactions can thus be directly deduced from the present results and those of our previous work (see Appendix~\ref{app:vision_cone}).

\subsection{Non reciprocal rank interaction}

We have also considered more general rank interactions, where the interaction parameter between particle $i$ and particle $j$ 
depends on their internal states, $\sigma_i(t)$ and $\sigma_j(t)$.
It is parameterized by a $2$ by $2$ matrix of couplings $\kappa_{\sigma,\sigma'}$ with $\sigma=\pm 1$ and $\sigma'=\pm 1$.
It is described by the equation of motion
\be \label{langevin1nonreciprocal}
\frac{dx_i}{dt} = \frac{1}{N} \sum_{j=1}^N \kappa_{\sigma_i(t),\sigma_j(t)}{\rm sgn}(x_i-x_j) - V'(x_i) + v_0 \sigma_i(t) \;.
\ee 

We have specialized to the interesting case of a non-reciprocal interaction such that
\be 
\kappa_{-+} = - \kappa_{+-} = b  \quad , \quad \kappa_{++} = \kappa_{--} = 0 \;.
\ee 
Restricting to $b>0$ by symmetry, this means that $+$ particles are attracted to $-$ particles, while $-$ particles are repelled by $+$ particles (and two particles with the same sign do not interact together). 
We find that for any $N \geq 2$ in the absence of external potential the particles eventually escape to infinity and there is no bound stationary state.

\begin{figure}
    \centering
    \includegraphics[width=0.49\linewidth, trim={0.3cm 2.2cm 0.3cm 2.4cm}, clip]{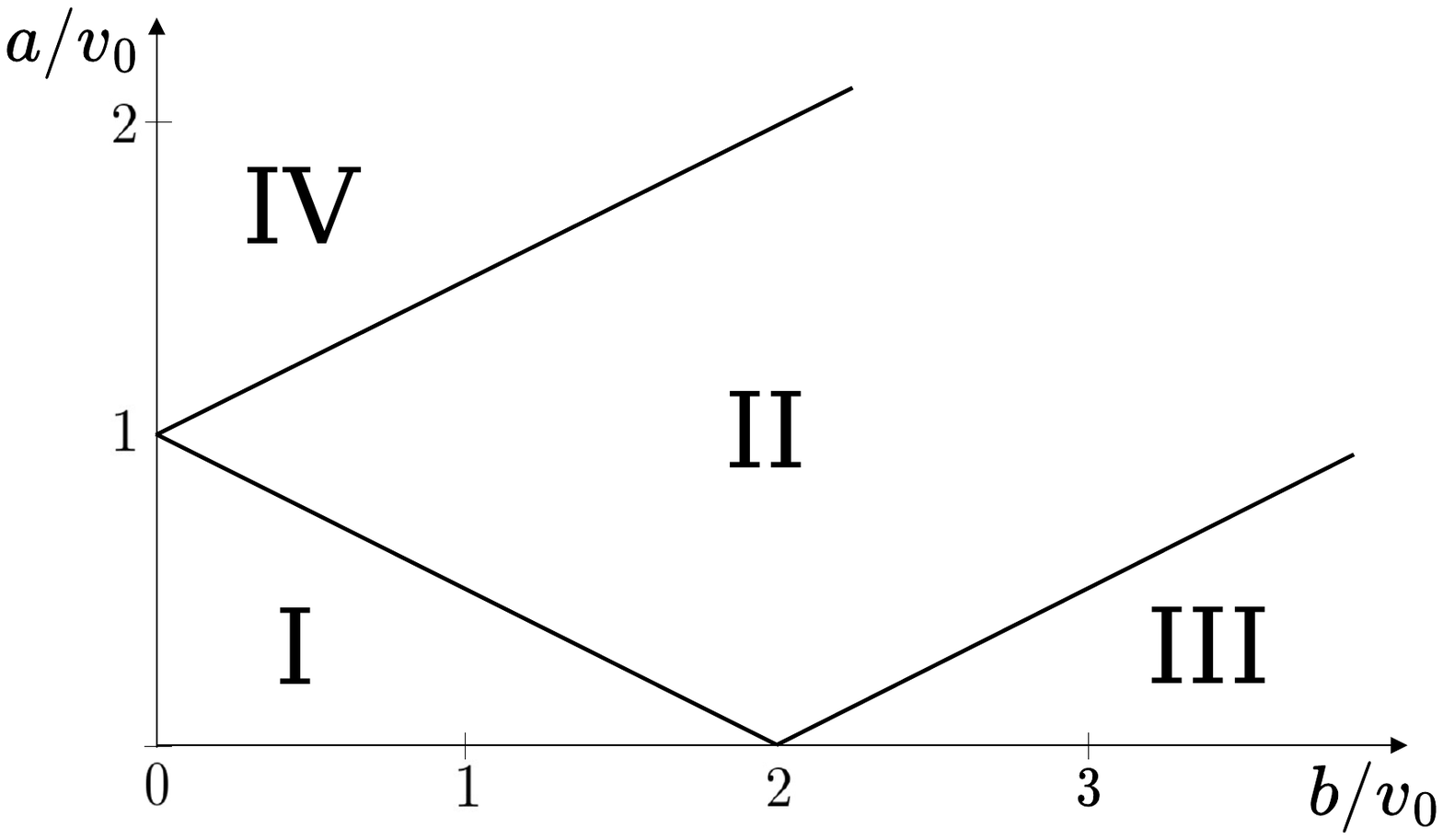}
    \includegraphics[width=0.49\linewidth, trim={0.4cm 2.2cm 0.4cm 2cm}, clip]{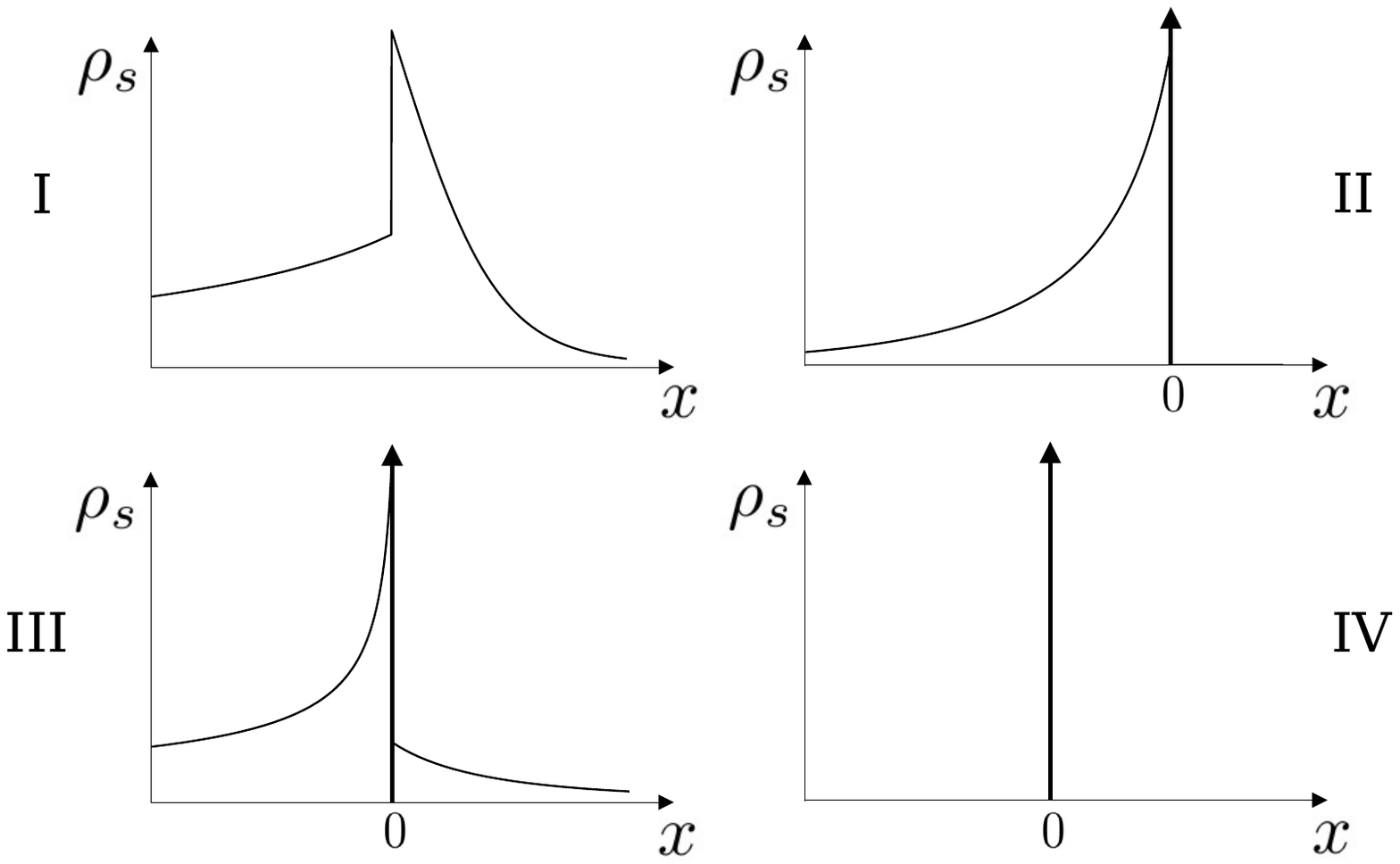}
    \caption{Left: Phase diagram of the non-reciprocal active rank diffusions. The phase diagram is
    symmetric upon $b \to - b$ and a parity transformation $x \to - x$. Right: the behavior of the total particle
    density in the four phases. The arrows denote delta function peaks in the density.}
    \label{phase_diagram_nonreciprocal}
\end{figure}

We have thus studied the system in presence of a confining linear external potential $V'(x)=a \, {\rm sgn}(x)$, with $a>0$. Our results are
summarized in Fig. \ref{phase_diagram_nonreciprocal}. We find that there are four phases. The phase IV corresponds to $a \geq v_0+b/2$ and is trivial: the particles cannot escape from $x=0$ and the density is thus a single delta peak at $x=0$.
In phases I, II and III the density is smooth outside of $x=0$ and 
decays exponentially as $\rho_s(x) \sim e^{- A_+ x} $ for $x\to+\infty$ and $\rho_s(x) \sim e^{A_- x}$ for $x\to-\infty$. The inverse sizes $A_\pm$ of the
bound state on the two sides are given below. The asymmetry of the stationary density is a result of the non-reciprocity of the interaction.
These three phases differ mainly by their behavior near and at $x=0$, 
and by the absence or presence of particles on either side of $x=0$. 
In phase I the density exhibits only a jump at $x=0$, and 
its support is the whole real line. In phases II and III the density exhibits a delta peak (i.e. a shock, a cluster of $+$ particles) at $x=0$. 
In phase II the density vanishes for $x>0$, i.e. all particles are either in the shock or at the left of $x=0$.
Surprisingly, in phase III the support of the density is again the whole real line. This phase
which exists for large non-reciprocity is dominated by fluctuations, see below. 
Let us now describe each phase in more details. 

In phase I, i.e. for $0<a<v_0-b/2$, the support of the density is the whole real line. The total density reads
\be \label{rhos_nr_intro}
\rho_s(x) = \begin{dcases} A_+ \frac{2 v_0-b}{2 b} \left(1 - \frac{1}{1 + W\left(\frac{b}{2 v_0} e^{- A_+ x + \frac{b}{2v_0}}\right) } \right) \quad \quad \quad \quad \ \ {\rm for} \ x>0 \;, \\
\rho_s(x) = A_- \frac{2 v_0+b}{2 b} \left(\frac{1}{1 + W\left(- \frac{b}{2 v_0} e^{A_- x - \frac{b}{2v_0}}\right) } -1 \right) \quad {\rm for} \ x<0 \;, \end{dcases} \quad {\rm with} \ A_\pm=\frac{8 a \gamma}{(2 v_0\mp b)^2- 4 a^2} \;,
\ee
where $W$ is the Lambert function, i.e. the real (and first) root of $W(z) e^{W(z)} = z$. Since $W(z) \simeq z$ at small $z$, 
the density decays 
exponentially at large $|x|$ with rates $A_\pm$. The decay is slower on the
negative side for $b>0$ since then $A_-<A_+$. 
The density exhibits a discontinuity at $x=0$ 
\be \label{ratiointro} 
\frac{ \rho_s(0^-) }{ \rho_s(0^+)} 
= \frac{1 - \frac{4 a^2}{(2 v_0 - b)^2}}{1 - \frac{4 a^2}{(2 v_0 + b)^2}} \;,
\ee 
which is smaller than unity for $b>0$. 

In phase II, i.e. for $v_0-b/2 < a < v_0+b/2$, the density $\rho_s(x)$ vanishes for $x>0$ and it has a delta peak at $x=0$ with weight
\be \label{weight2intro}
\frac{1}{2}-r(0^-)= \frac{2 a}{2 a + b + 2 v_0} \;,
\ee 
containing only $+$ particles. For $x<0$ it is given by
\be \label{rhos_nr_neg2_intro}
\rho_s(x) = A_- \frac{2 v_0+b}{2 b} \left(\frac{1}{1 + W\left(- \frac{2b}{2a+b+2 v_0} e^{A_- x - \frac{2b}{2a+b+2v_0}}\right) } -1 \right) \;,
\ee
where $A_-$ is still given by \eqref{rhos_nr_intro}. One has $\rho_s(0^-)=(b+2 v_0) A_-/(2 a - b + 2 v_0)$. 

Note that $A_+$ diverges as the phase II is approached from phase I, and all $+$ particles
for $x \geq 0$ become part of the cluster at $x=0$, i.e. the weight \eqref{weight2intro} is non-zero
on the frontier between phase I and phase II 
and $\rho_s(0^+)$ diverges in that limit.

Finally, in phase III, i.e. for $a < v_0-b/2$, the density $\rho_s(x)$ is again non zero both for $x>0$ and
$x<0$ and it has a delta peak at $x=0$ of weight
\be \label{weight3intro} 
r(0^+)-r(0^-)=\frac{2 a \left(b^2-2 a b-4v_0^2\right)}{b \left(b^2-4a^2-4 v_0^2\right)}
\ee
containing only $+$ particles. For $x \neq 0$ the density is given by
\be \label{rhos_shock_intro}
\rho_s(x) = \begin{dcases} A_+ \frac{b-2 v_0}{2 b} \left(\frac{1}{1 + W\left(-B_+ e^{- A_+ x - B_+}\right) } - 1 \right) \quad , \quad \quad \quad \quad  x>0 \;, \\
\rho_s(x) = A_- \frac{2 v_0+b}{2 b} \left(\frac{1}{1 + W\left(- B_- e^{A_- x -B_-}\right) } -1 \right) \quad , \quad x<0 \;,  \end{dcases} \quad B_\pm = \frac{(b\pm 2v_0)(b-2a\mp2v_0)}{b^2-4a^2-4v_0^2} \;, 
\ee
where $A_\pm$ are the same as in \eqref{rhos_nr_intro}.

Note again that $A_+$ diverges as the phase II is approached from phase III. However in this case the fraction of particles on the right side of $x=0$ vanishes continuously as one approaches this line, i.e. $\frac{1}{2}-r(0^+)\to 0$. Thus the weight of the delta peaks in the two phases, given by \eqref{weight2intro} and \eqref{weight3intro} respectively, match on the frontier between phase III and II (one finds $r(0^+)-r(0^-)=\frac{1}{2} - \frac{v_0}{b}$ in both cases).
\\

\begin{figure}[t]
    \centering
    \includegraphics[width=0.32\linewidth]{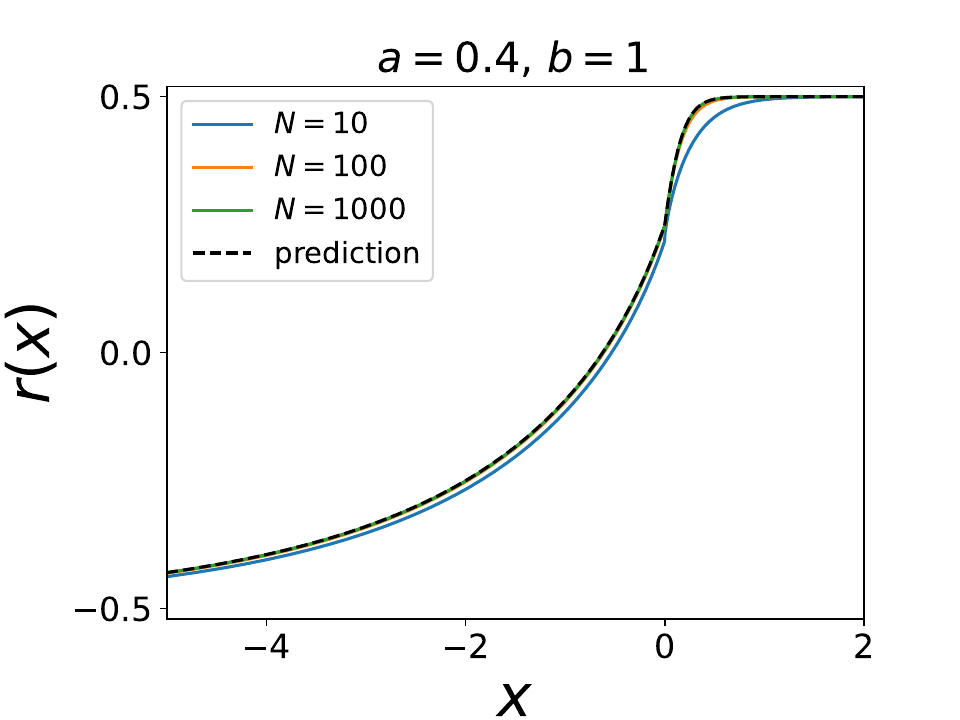}
    \includegraphics[width=0.32\linewidth]{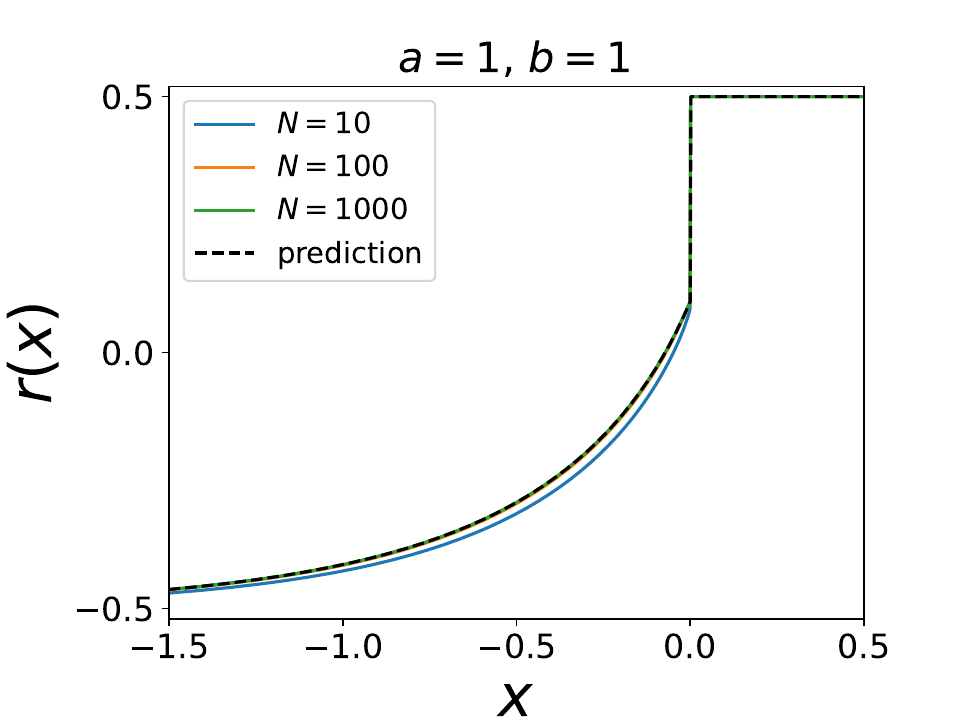}
    \includegraphics[width=0.32\linewidth]{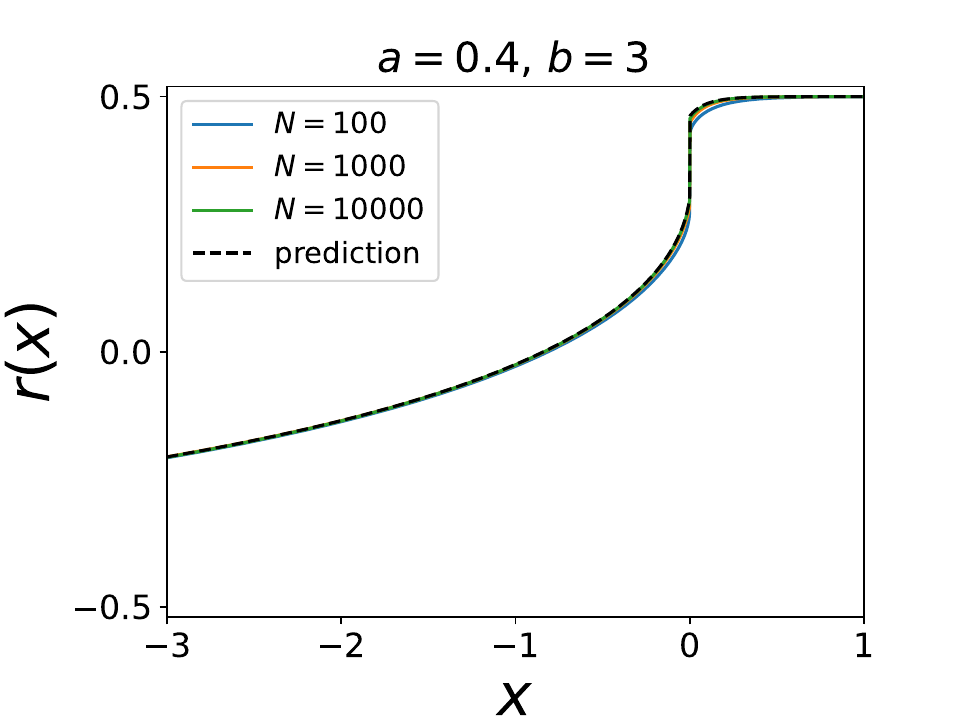}
    \includegraphics[width=0.32\linewidth]{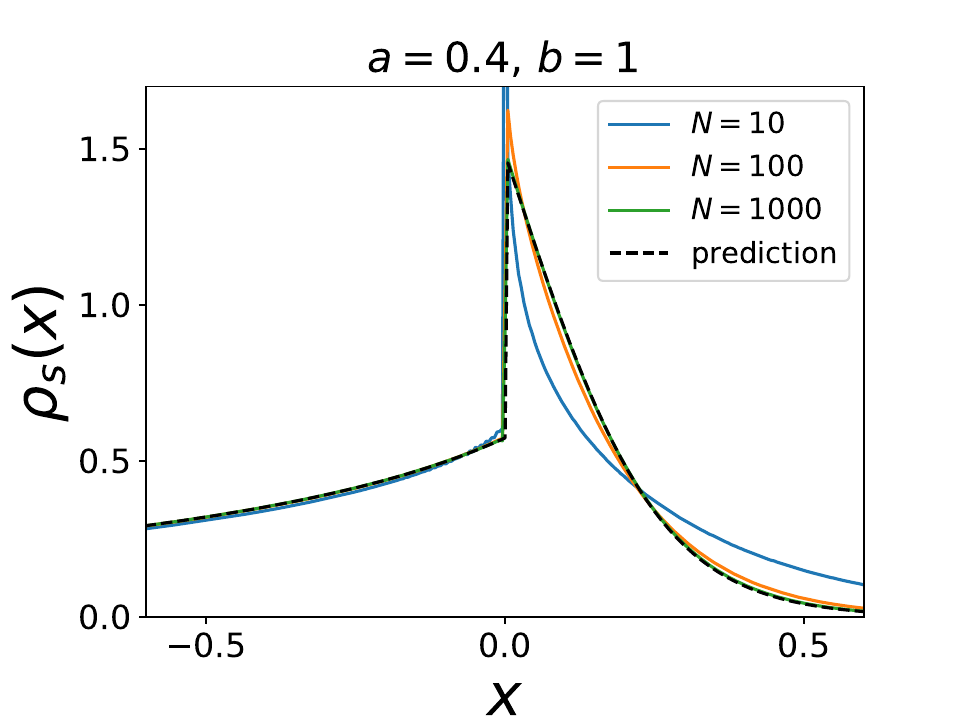}
    \includegraphics[width=0.32\linewidth]{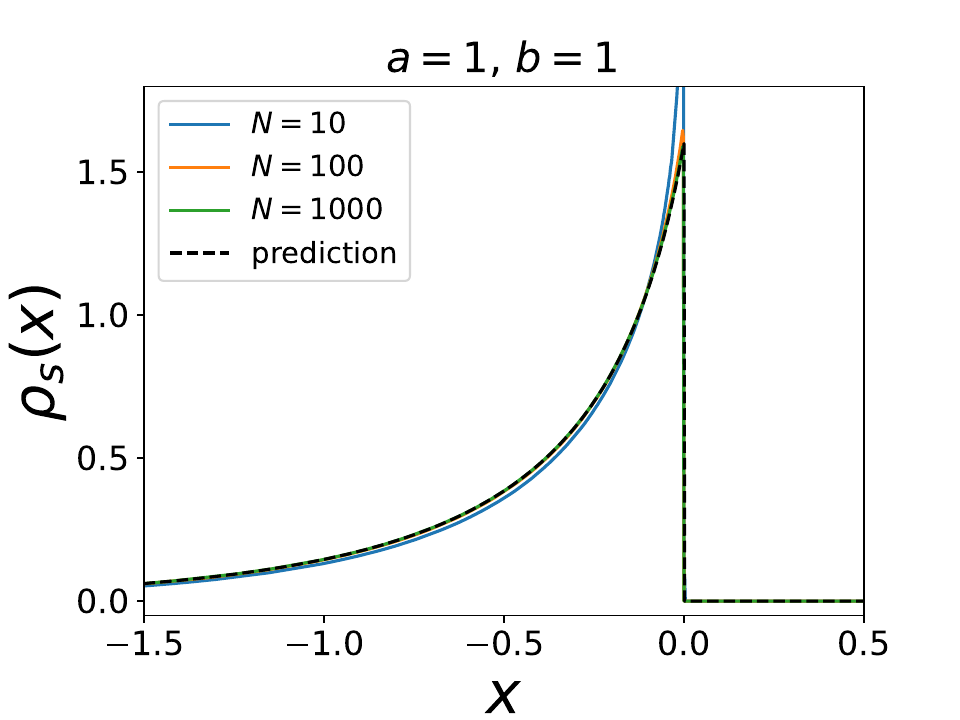}
    \includegraphics[width=0.32\linewidth]{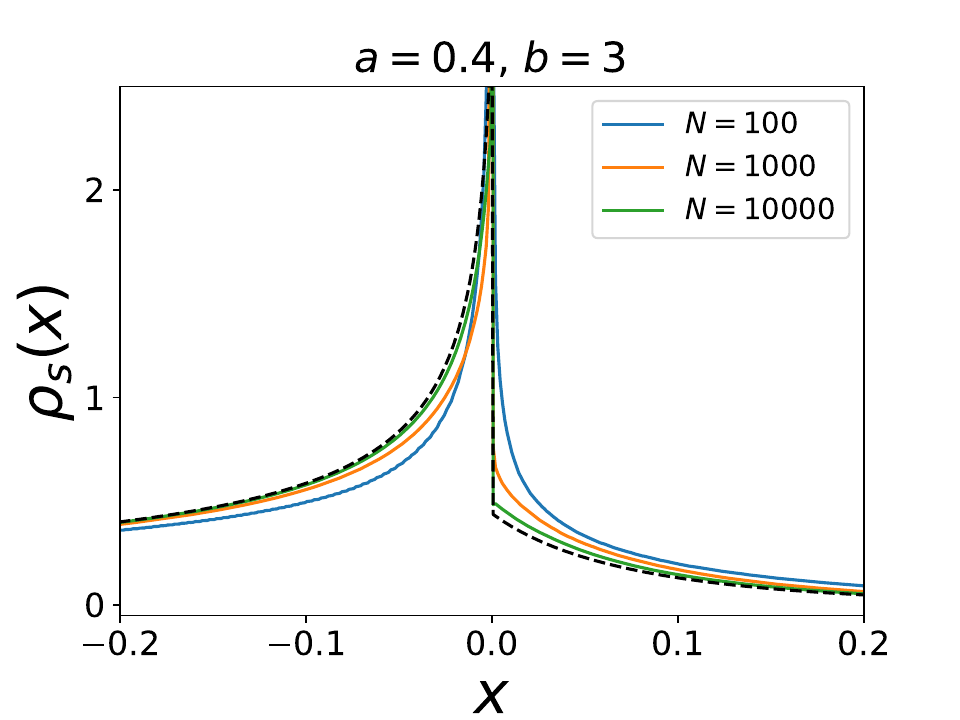}
    \caption{Rank field $r(x)$ (top) and total density $\rho_s(x)=r'(x)$ (bottom) obtained from simulations with $\gamma=1$ and $v_0=1$, for increasing values of $N$. The delta peaks are not shown on the densities for the phases II and III, but they are visible as discontinuities in $r(x)$. Left: $a=0.4$ and $b=1$ (phase I). The dashed black line corresponds to the prediction for infinite $N$ \eqref{rhos_nr_intro}. Note the discontinuity of the density at $x=0$, with
    however no shock (no delta peak). Center: $a=1$ and $b=1$ (phase II). The dashed black line corresponds to \eqref{rhos_nr_neg2_intro}. The density is zero for $x>0$ for any $N$ (with a delta peak at $x=0$). Right: $a=0.4$ and $b=3$ (phase III) the dashed black line corresponds to \eqref{rhos_shock_intro}. Larger values of $N$ are used here since the convergence in $N$ is slower than in the other phases.}
    \label{plots_nonreciprocal}
\end{figure}

These results can be interpreted by considering the force seen by the particles near $x=0$, due to the combined effect of the potential and the non-reciprocal attractive interaction.
In phase I, one finds that all the $+$ particles move towards the right and the $-$ particles towards the left, until they change sign. The total force felt by the rightmost particle, assuming that it is a $+$ particle, is $v_0-a-b/2$, which is indeed positive inside phase I. When this force becomes negative, this marks the transition to phase II. In this phase, all the $+$ particles are attracted towards $x=0$, while the $-$ particles still move towards the left. This explains why a cluster of $+$ particles forms at $x=0$ and no particle can access $x>0$. The force felt by the leftmost particle, assuming that it is a $-$ particle, is $-v_0+a-b/2$. When this becomes positive, this marks the transition to phase IV where all particles are attracted towards $x=0$ indifferently of their sign and remain there. The behaviour of the particles in phase III is less intuitive. In this phase, all the $+$ particles are attracted towards $x=0$ while the $-$ particles are repelled either towards $x>0$ or $x<0$. When a $+$ particle from the cluster at $x=0$ becomes a $-$ particle, the side towards which it is directed is determined by the fluctuations. Thus this phase is more sensitive to the fluctuations than the other phases.

The analytical results and the qualitative description above are valid in the limit$N\to+\infty$. In Fig.~\ref{plots_nonreciprocal} we compare our analytical predictions for $r(x)$ and $\rho_s(x)$ for $N\to +\infty$ to numerical results obtained through simulations of the stochastic dynamics of equation \eqref{langevin1nonreciprocal} and averaging over time, in the stationary state,
for finite values of $N$. We observe that the 4 phases already exist for finite $N$, with the same qualitative behaviours (presence of a true delta peak at $x=0$, absence or presence of a smooth density of particles on the right...). The quantitative values differ from our predictions for small values of $N$, but rapidly converge as $N\to+\infty$. In phase IV the convergence is slower due to the peculiar role played by the fluctuations in this phase, as discussed above.

\subsection{Outline of the paper}

In the rest of the paper we give the derivations for all the results presented here. 
In Section \ref{sec:derivation} we study active rank diffusions in a harmonic trap. We start by deriving the general equations for the model
and we obtain a parametric representation of the stationary densities in the large $N$ limit.
After recalling the non-interacting case, we first study the 
the special line $\mu=2\gamma$ in the phase diagram for which a more explicit solution can be obtained. Then we consider the general case
and use the previously obtained representation to study the presence or absence of shocks and the
edge behavior of the densities. 
We finally discuss in detail the phase diagram, and conclude with a study of the limit of small $\gamma$.
In Section \ref{sec:nonreciprocal} we consider the case of RTP's with non-reciprocal rank interactions. 
We first derive the general equations for arbitrary non-reciprocal rank interactions between the $\pm$ states. We specialize 
to a simple version of the model presented above and derive the exact densities in the large $N$ limit. 
We then examine the three non-trivial phases in detail and obtain the phase diagram.


\section{ Active rank diffusions in a harmonic trap} \label{sec:derivation}

\subsection{General equations for the rank fields}

Using the Dean-Kawasaki approach  \cite{Dean,Kawa}, one can establish starting from \eqref{langevin1} 
, as in \cite{PLDRankedDiffusion} and \cite{TouzoDBM2023}, an exact stochastic evolution equation
for the density fields, which takes the following form 
\be \label{eqrho1}
\partial_t \rho_\sigma  = T \partial_x^2 \rho_\sigma  +  \partial_x \rho_\sigma  \big( - v_0 \sigma +  V'(x) + \bar \kappa \int dy \, \rho_s(y,t) {\rm sgn}(x-y) \big) + \gamma \rho_{- \sigma} - \gamma \rho_{\sigma}
+ O(\frac{1}{\sqrt{N}}) \;,
\ee
where the $O(\frac{1}{\sqrt{N}})$ term represents the passive and active noises,
see Supp. Mat. in \cite{activeRDshort}, and Sec. III A of \cite{TouzoDBM2023}.
It is convenient to define, as in \cite{PLDRankedDiffusion,activeRDshort},
the rank fields
\be \label{rankdef1}
r(x,t) = \int^x_{-\infty} dy \, \rho_s(y,t)\, -\, \frac{1}{2} \quad , \quad s(x,t) = \int_{-\infty}^x dy \, \rho_d(y,t) \; .
\ee
Focusing from now on on the large $N$ limit, and henceforth neglecting the noise terms in \eqref{eqrho1},
one obtains from \eqref{eqrho1} and \eqref{rankdef1} two coupled deterministic differential equations (see \cite{activeRDshort})
\bea \label{system1}
\!\!\!\!\!\!  \partial_t r &=& T \partial_x^2 r - v_0 \partial_x s
+ 2 \bar \kappa r \partial_x r + V'(x) \partial_x r \; , \\
\!\!\!\!\!\!  \partial_t s &=& T  \partial_x^2 s  - v_0 \partial_x r  + 2  \bar \kappa  r \partial_x s + V'(x) \partial_x s - 2 \gamma s \label{system2} \;.
\eea 
These equations are valid both for the attractive and the repulsive case. Since $\rho_s(x,t)$ is positive and normalized to 1, $r(x,t)$ must be an increasing function with $r(\pm \infty,t)=\pm \frac{1}{2}$. One sees that $\partial_t s(+\infty,t)=-2 \gamma s(+\infty,t)$, hence at large time
$s(\pm \infty,t)=0$. 
In the passive case $v_0=0$ the first equation recovers Burger's equation which
describes usual rank diffusion \cite{PLDRankedDiffusion}. 
One generally expects that in the diffusive limit $v_0,\gamma \to +\infty$ with 
$T_a=\frac{v_0^2}{2 \gamma}$ fixed, RTP's behave as Brownian particles. This also holds here. Indeed, in \eqref{system2} only two terms 
remain relevant in that limit, and one obtains $s \simeq - \frac{v_0}{2\gamma} \partial_x r$. Inserting in \eqref{system1} the active term 
$- v_0 \partial_x s(x,t)$ becomes $T_a \partial_x^2 r(x,t)$, i.e. a diffusive thermal term with effective temperature $T_a$. 
\\

In the remainder of the paper we will only consider RTP noise, setting $T=0$.

\subsection{A parametric representation for the stationary state in a general confining potential}

We start from the stationary Dean-Kawasaki equation (i.e. with time derivatives set to zero) for an arbitrary confining external potential $V(x)$,
\bea \label{stationary_eq}
&& 0 = - v_0 s'  - 2 \kappa \, r r'  + V'(x) r' \;, \\
&& 0 = - v_0 r' - 2 \kappa r s' - 2 \gamma s + V'(x) s' \nonumber
\eea 
(where the primes now denote spatial derivatives), which must be solved with boundary conditions $r(\pm \infty)= \pm \frac{1}{2}$ and
$s(\pm \infty)=0$. Let us recall that $\kappa>0$ corresponds to repulsive interaction,
while $\kappa=- \bar \kappa <0$ corresponds to attractive interaction.
The function $r(x)$ is increasing, but may have plateaus or jumps (shocks)
at some positions. On each interval where $r(x)$ is strictly monotonous and smooth, 
it is always possible to write 
\be 
V'(x) = W'(r(x)) 
\ee 
and hope to determine $V(x)$ from $W(r)$ later on. Integrating the first equation one obtains 
\be 
s = \frac{1}{v_0} (- \kappa (r^2 - \frac{1}{4}) + W(r) ) \;, \label{srel1}
\ee 
where $W(r)$ should satisfy some boundary conditions, including $W(\pm\frac{1}{2})=0$. We introduce
\be 
U(r) = - \kappa (\frac{1}{4} - r^2) - W(r) \quad , \quad s = -U(r)/v_0 \;. \quad , \quad U(\pm \frac{1}{2}) = 0 \;. \label{defU}
\ee 
The second equation in \eqref{stationary_eq} becomes
\be
(v_0^2 - U'(r)^2)r' = 2\gamma U(r) \;.
\ee
Integrating this equation, and restricting for convenience to potentials $V(x)$ which are
even in $x$, such that $r(x)$ is odd, and $r(0)=0$ (assuming that there is no
shock at $x=0$) one finds an equivalent parametric representation of Eqs. \eqref{stationary_eq}
in the form
\bea
&& x = \int_{0}^{r(x)} dr \frac{v_0^2 - U'(r)^2}{2 \gamma U(r) } \;, \label{xparamU} \\
&& - V'(x) = -2 \kappa r + U'(r) \;, \label{Ueq1}
\eea
together with $s = -U(r)/v_0$. It is parametric in the sense that for any
choice of $U(r)$ one obtains in principle a solution $(r(x),s(x))$ of
Eqs. \eqref{stationary_eq} for "some" potential $V(x)$ determined a posteriori.

\subsection{A general class of solutions for the harmonic potential}

We now specialize to a harmonic external potential $V'(x)=\mu x$. 
In this case if the solution is unique, $r(x)$ is odd, and $s(x)$ is even in $x$.
We apply the method of the previous subsection.
Eq. \eqref{Ueq1} becomes
\be \label{U_harmonic}
\mu x = 2 \kappa r - U'(r) \, .
\ee
We will first determine $U(r)$, which will allow for finding $r(x)$ through the above equation.
To this aim, let us 
replace $x$ using \eqref{xparamU} and take the derivative with respect to $r$. We obtain
\be
U''(r) - 2 \kappa + \mu \frac{v_0^2 - U'(r)^2}{2\gamma U(r)} = 0 \, ,
\ee
which can be rewritten as
\be \label{eqU} 
U'(r)^2 - v_0^2 = 2a U(r)U''(r) - b U(r) \quad \text{with } a=\frac{\gamma}{\mu} \text{ and } b = 4a\kappa \, .
\ee
This equation can be solved parametrically by introducing the function $g(u)$ such that $g(U(r))=U'(r)$. This implies that $U''=g'(U)U'=g'(U)g(U)$, and thus $g(u)$ satisfies
\be
g^2(u)-v_0^2 = 2aug(u)g'(u)-bu \, .
\ee
Introducing $G(u)=g^2(u)$ we finally obtain
\be
a u \, G'(u) - G(u) - bu + v_0^2 = 0 \;, \label{eqG1}
\ee
for which the solutions are
\be \label{G_expression}
G(u) = \begin{dcases} C u^{1/a} + \frac{b}{a-1}u + v_0^2 \quad \text{for } a \neq 1 \;, \\
C u + b u\ln u + v_0^2 \quad \quad \ \ \text{for } a = 1 \;, \end{dcases}
\ee
with $C$ a constant to be determined.

We now have to solve $(U')^2=G(U)$. Hence we need to know the sign of $U'(r)$. Since from \eqref{defU} we have that $s=-U(r)/v_0$,
and we expect that in the stationary state there is a surplus of $-$ particles for $x<0$ and $+$ particles for $x>0$. Hence $s(x)$ should be negative, with $s(\pm \infty)=0$ and a minimum at $x=0$. Since $r(x)$ is an increasing function with $r(0)=0$, this suggests that $U(r)$ should have a maximum at $r=0$. Hence $U'(r)$ should be positive for $r<0$ and negative for $r>0$, with $U'(0)=0$ (the last point can also be deduced from \eqref{Ueq1} with $V'(x)=\mu x$, since $r(0)=0$ by symmetry). This leads to 
\be \label{signUprime}
U'=-{\rm sgn}(r)\sqrt{G(U)}  \quad , \quad U'(0)=0 \;.
\ee 
The second property will allow to determine the constant $C$ (see below). 
Since $r(x)$ is an odd function of $x$, from now on we will focus on $r>0$ and $x>0$. 
Using the boundary condition in \eqref{defU}, $U(\frac{1}{2})=0$, 
we thus obtain a parametric equation for $U(r)$, which together with \eqref{U_harmonic},
also determines $r(x)$
\bea \label{eqU_G}
&& \frac{1}{2} - r = \int_{0}^{U(r)} \frac{du}{\sqrt{G(u)}} \, , \\
&& \mu x = 2 \kappa r - U'(r) \;, \label{eq_xU}
\eea
where $G(U)$ is given in \eqref{G_expression}. 
The constant $C$ is fixed by the condition $G(U(0))=0$, i.e. denoting $u_0=U(0)$,
\be \label{C_expression}
C = \begin{dcases} -\frac{b}{a-1}u_0^{1-\frac{1}{a}} - v_0^2 u_0^{-1/a} \quad \text{for } a \neq 1 \;, \\
-b\ln u_0 - \frac{v_0^2}{u_0} \quad\quad \quad \quad \quad \ \text{for } a = 1 \;. \end{dcases}
\ee
where $u_0$ is the solution of
\be \label{eq_u0}
\frac{1}{2} = I(u_0) \quad , \quad  I(u_0) := \begin{dcases} u_0 \int_0^{1} \frac{dw}{\sqrt{ v_0^2 (1 - w^{1/a}) + \frac{b u_0}{a-1} (w - w^{1/a}) }} \quad \text{for } a \neq 1 \;, \\
u_0 \int_0^{1} \frac{dw}{\sqrt{ v_0^2 (1 - w) + b u_0 w \ln w }} \quad \quad \quad \quad \ \ \text{for } a = 1 \;, \end{dcases}
\ee
which has been obtained taking $r=0$ in \eqref{eqU_G} and noting that one 
can rewrite, with $u_0=U(0)$,
\be
G(u) = \begin{dcases} v_0^2 \left(1 - \left(\frac{u}{u_0}\right)^{1/a}\right) + \frac{b u_0}{a-1} \left(\frac{u}{u_0} - \left(\frac{u}{u_0}\right)^{1/a}\right)  \quad \text{for } a \neq 1 \;, \\
v_0^2 \left( 1-\frac{u}{u_0} \right) + { bu \ln \frac{u}{u_0} \hspace{4cm} \text{for } a = 1 \;.} \end{dcases}
\ee
In the repulsive case, $b>0$, an analysis of the argument of the square root in \eqref{eq_u0} shows that 
for any $a>0$, $I(u_0)$ is a strictly increasing function 
of $u_0 \in [0,u_0^{\rm max}]$, which diverges as $u_0 \to u_0^{\rm max}=v_0^2/b$, and evaluates to $0$ at $u_0=0$. In the attractive case, $b<0$, one has the same result with $u_0^{\rm max}=+\infty$.
Thus there always exists a unique value of $u_0>0$ for which \eqref{eq_u0} is satisfied.
Once $G(u)$ is known, this leads to a parametric representation for $r(x)$ obtained by varying $U \in [0,u_0]$
\bea \label{param_r}
&&  r = \frac{1}{2} - \int_{0}^{U} \frac{du}{\sqrt{G(u)}}  \, , \\
&& \mu x = \kappa  - 2 \kappa  \int_{0}^{U} \frac{du}{\sqrt{G(u)}} + \sqrt{G(U)} \;,
\eea
where $G(U)$ is given in \eqref{G_expression}.
One can also obtain a similar representation for $\rho_s(x)$. Taking the derivative of \eqref{U_harmonic} w.r.t. $x$ yields
\be 
 \rho_s(x)= \frac{\mu}{2\kappa - U''(r)} \;,
\ee
where, using \eqref{signUprime}, one has
\be \label{Usecond}
U''(r) = -\frac{U'(r) G'(U(r))}{2\sqrt{G(U(r))}} = \frac{1}{2} G'(U(r)) \;,
\ee
with 
\be
G'(u) = \begin{dcases} \frac{C}{a} u^{\frac{1}{a}-1} + \frac{b}{a-1}  = - \frac{1}{a} \left( \frac{v_0^2}{u_0} + \frac{b}{a-1} \right) \left( \frac{u}{u_0} \right)^{\frac{1}{a}-1} + \frac{b}{a-1}  \quad \text{for } a \neq 1 \;, \\
C + b (1+\ln u)  = b\ln\frac{u}{u_0} + b-\frac{v_0^2}{u_0} \hspace{3.9cm} \text{for } a = 1 \;. \end{dcases}
\ee
To obtain a parametric representation of the density one thus should again vary $U \in [0,u_0]$ and compute
\bea \label{param_rho}
&& \rho_s = \frac{\mu}{2\kappa - \frac{1}{2} G'(U)} \, , \\
&& \mu x = \kappa  - 2 \kappa  \int_{0}^{U} \frac{du}{\sqrt{G(u)}} + \sqrt{G(U)} \;.
\eea

Note that all the results derived in this subsection are valid both for repulsive ($\kappa>0$) and
attractive ($\kappa<0$) interactions, as long as there are no shocks. The condition for the
absence of shocks will be studied below. We will now examine this general solution in more
details, starting with the non interacting case $\kappa \to 0$, then the 
the repulsive case, and finally the attractive case (where the shocks can occur). 

\subsection{Non-interacting case $\kappa=0$}

For $\kappa=0$, the above formulas recover the known result for the density of a stationary RTP in a harmonic potential (see e.g. \cite{DKM19}). Indeed in this case one has $b=0$, and thus
\be \label{int_noninteracting}
\frac{1}{2} - r = \int_0^{U(r)} \frac{du}{v_0 \sqrt{1- (\frac{u}{u_0})^{1/a}}} = \frac{U(r)}{v_0} \, _2F_1\left(\frac{1}{2}, a;a+1; \left(\frac{U(r)}{u_0} \right)^{1/a}\right) \;.
\ee 
The equation \eqref{eq_u0} for $u_0$ is solved by  
\be
u_0= \frac{v_0}{2 \sqrt{\pi}} \frac{ \Gamma(a+\frac{1}{2})}{\Gamma(a+1)} \;.
\ee
One the other hand one has
\be 
\mu x = - U'(r) = \sqrt{G(U(r))}= v_0 \sqrt{1 - \left(\frac{U(r)}{u_0}\right)^{1/a} } \;,
\ee 
which leads to
\be 
U(r(x)) = u_0 \left(1- \left(\frac{\mu x}{v_0}\right)^2 \right)^a \;.
\ee 
This implies
\be 
\frac{1}{2} - r = \frac{u_0}{v_0} \left(1 - \left(\frac{\mu x}{v_0}\right)^2\right)^a \,
   _2F_1\left(\frac{1}{2},a;a+1; 1- \left(\frac{\mu x}{v_0}\right)^2 \right) \;.
\ee 
Taking a derivative and replacing $a=\frac{\gamma}{\mu}$, one finally obtains
\be 
\rho_s(x) = \frac{2 \gamma u_0}{v_0^2} \left(1 - \left(\frac{\mu x}{v_0}\right)^2\right)^{\frac{\gamma}{\mu}-1} 
= \frac{2}{4^{\gamma/\mu} B(\frac{\gamma}{\mu},\frac{\gamma}{\mu})} \frac{\mu}{v_0} \left(1 - \left(\frac{\mu x}{v_0}\right)^2\right)^{\frac{\gamma}{\mu}-1} \;,
\ee 
where $B(\alpha, \beta)= \Gamma(\alpha) \Gamma(\beta)/\Gamma(\alpha+\beta)$ is the beta function, and
we have used the identity $B(\alpha,\alpha)= 2^{1-2 \alpha} B(\frac{1}{2},\alpha)$.
Using $s=-U(r)/v_0$ and taking the derivative w.r.t $x$ one also finds
\be
\rho_d(x) = \frac{\mu x}{v_0} \rho_s(x) \;,
\ee
which leads to
\be
\rho_\pm(x) = \frac{1}{4^{\gamma/\mu} B(\frac{\gamma}{\mu},\frac{\gamma}{\mu})} \frac{\mu}{v_0} \left(1 \pm \frac{\mu x}{v_0}\right)^{\frac{\gamma}{\mu}} \left(1 \mp \frac{\mu x}{v_0}\right)^{\frac{\gamma}{\mu}-1} \;.
\ee

\subsection{Special case $a=1/2$, i.e. $\mu=2 \gamma$}  \label{sec:aspecial}

We start with the case $a=1/2$ which can be solved in a more explicit form (in that case $C$ is dimensionless). 
We first present the solution assuming that there are no shocks. 
These solutions will be fully valid in the repulsive case $\kappa>0$. In the attractive case $\kappa<0$ we find
that there are always shocks, however the above solution is still valid in some range of values of $r$ or $x$.

\subsubsection{Solution in the absence of shocks and repulsive case $\kappa>0$}

We start by determining the function $U(r)$. 
In the particular case $a=1/2$, the integral in \eqref{eqU_G} can be computed, yielding
\bea \label{3cases} 
\frac{1}{2} - r &=& \left[ -\frac{1}{\sqrt{C}} \, {\rm arcsinh} \big( \sqrt{\frac{C^2}{v_0^2 C-b^2}} (\frac{b}{C}-u) \big) \right]_0^{U(r)} \quad \ \text{if } C>\frac{b^2}{v_0^2} \, ,  \\
\frac{1}{2} - r &=& \left[ -\frac{1}{\sqrt{C}} \, {\rm arcosh} \big( \sqrt{\frac{C^2}{b^2-v_0^2 C}} (\frac{b}{C}-u) \big) \right]_0^{U(r)} \quad \ \ \text{if } 0<C<\frac{b^2}{v_0^2} \, , \nonumber \\
\frac{1}{2} - r &=& \left[ -\frac{1}{\sqrt{|C|}} \arcsin \big( \sqrt{\frac{C^2}{b^2-v_0^2C}} (\frac{b}{C}-u) \big) \right]_0^{U(r)} \quad \text{if } C<0 \, , \nonumber \\
\text{with } C&=&\frac{2b}{u_0} - \frac{v_0^2}{u_0^2} \, , \nonumber
\eea
where we recall that $u_0=U(0)$ is solution of \eqref{eq_u0}, although we do not need to determine it here (see below).
Note that in this case one has $b=2\kappa$. For now the sign of $b$ is still arbitrary.

Let us consider first the subcase $C>0$, which corresponds to the first two equations in \eqref{3cases}.
Inverting the relations \eqref{3cases}, and denoting $C=c^2$ ($c>0$) one obtains
\be \label{U_cpos}
U(r) = \frac{1}{c^2} \left(  2\kappa \left(1- \cosh(c (\frac{1}{2} - r) ) \right)  + c v_0 \sinh(c (\frac{1}{2} - r) ) \right) \;.
\ee
One can check that it satisfies $U(1/2)=0$ and that it obeys \eqref{eqU}. The condition $G(U(0))=0$ is then equivalent to
\be \label{tanh} 
\frac{\tanh(c/2)}{c/2}= \frac{v_0}{\kappa} \;.
\ee 
which determines $c$. It has a solution only for $\kappa> v_0$, 
in which case 
$c$ takes values from $c=0$ for $\kappa=v_0$ to $c\to+\infty$ for $v_0/\kappa=0$. 

Let us now discuss the subcase $C<0$, which corresponds to $\kappa < v_0$. Then one should use the third equation \eqref{3cases} with $C<0$. Denoting now $C=- c^2$ ($c>0$), we obtain
\be \label{U_cneg}
U(r) = \frac{1}{c^2} \left(  - 2\kappa \left(1- \cos(c (\frac{1}{2} - r) ) \right)  + c v_0 \sin(c (\frac{1}{2} - r) ) \right) \;,
\ee
and $G(U(0))=0$ is now equivalent to
\be \label{tan} 
\frac{\tan(c/2)}{c/2}= \frac{v_0}{\kappa} \;,
\ee 
in which case $c$ varies from $c=0$ for $\kappa=v_0$ to $c=\pi$ for $v_0/\kappa\to+\infty$. 

We now determine the rank field $r(x)$. We start with the special case $\kappa=v_0$, which corresponds to 
$c=0$. Taking the limit, one obtains
\be 
U(r) = -\frac{1}{8} (2 r-1) (2\kappa (2 r-1)+4 v_0) = v_0(\frac{1}{4}-r^2) \;.
\ee 
This quadratic function is a solution of \eqref{eqU} for any $\kappa$ and $v_0$, but it satisfies $U'(0)=0$ only for $\kappa=v_0$. In this case \eqref{eq_xU} simply becomes
\be 
\mu x = 4\kappa r \;.
\ee 
Thus we obtain, for $\mu=2\gamma$ and $\kappa=v_0$, the very simple solution
\be \label{r_square_ahalf}
r(x) = \frac{\mu x}{4\kappa} \quad \text{for } x\in[-\frac{2\kappa}{\mu},\frac{2\kappa}{\mu}] \;, 
\ee
i.e. the total density $\rho_s(x)=\frac{\mu}{4 \kappa}$ is constant on the support $[-\frac{2\kappa}{\mu},\frac{2\kappa}{\mu}]$. Then one has 
\be \label{s_square_ahalf}
s(x) = -\frac{U(r(x))}{v_0} = r(x)^2-\frac{1}{4} = \frac{\mu^2 x^2}{16\kappa^2}- \frac{1}{4} \;,
\ee
and thus $\rho_d(x)=\frac{\mu^2 x}{8\kappa^2}$ for $x\in[-\frac{2\kappa}{\mu},\frac{2\kappa}{\mu}]$.

Let us now consider the case where $\kappa<v_0$. We substitute $\kappa=\frac{v_0 c}{2\tan(c/2)}$, from \eqref{tan}. Then the solution \eqref{U_cneg} reads
\be 
U(r) = \frac{v_0}{c^2} \left(  - \frac{c}{\tan(c/2)} \left(1- \cos(c (\frac{1}{2} - r) ) \right)  + c  \sin(c (\frac{1}{2} - r) ) \right) \;,
\ee
which satisfies $G(U(0))=0$. The rank field is obtained by inverting
\bea \label{x_cneg1}
\mu x = 2 \kappa r -U'(r) = 2 \kappa r + v_0 \frac{\sin(c r)}{\sin(c/2)} \;.
\eea
Replacing again $\kappa=\frac{v_0 c}{2\tan(c/2)}$ in \eqref{x_cneg1} 
one obtains $r(x)$ by inverting
\be \label{r_repulsive1}
\frac{\mu x}{v_0} \sin(c/2)  = g_c(r) \quad , \quad g_c(r) :=\sin (cr) + cr \cos(\frac{c}{2}) \quad , \quad \frac{\tan(c/2)}{c/2}= \frac{v_0}{\kappa} \;,
\ee 
where now $c$ is the only parameter (up to a rescaling of $x$). 

From now on we restrict to the repulsive case $\kappa>0$, so that $c$ varies between $0$ and $\pi$.
For $c\in[0,\pi]$, $g_c(r)$ is an increasing function on $[0,1/2]$, so that this relation can be inverted without any issue (i.e. there is no shock). The density has a finite support $[-x_e,x_e]$ where the value of the edge $x_e$ is obtained by taking $r=1/2$ in \eqref{x_cneg1},
\be 
x_e = \frac{v_0 + \kappa}{\mu} \;. \label{edge0}
\ee 

For $\kappa>v_0$, one has
\bea 
\mu x = 2 \kappa r -U'(r) = 2 \kappa r + v_0 \frac{\sinh(c r)}{\sinh(c/2)} \quad , \quad \frac{\tanh(c/2)}{c/2}= \frac{v_0}{\kappa} \;,
\eea 
or equivalently $r(x)$ is obtained by inverting
\be \label{r_repulsive2}
\frac{\mu x}{v_0} \sinh(c/2)  = \tilde g_c(r) \quad , \quad \tilde g_c(r)=\sinh (cr) + cr \cosh(\frac{c}{2}) \quad , \quad \frac{\tanh(c/2)}{c/2}= \frac{v_0}{\kappa} \;,
\ee 
where $c\in[0,+\infty]$. Again $\tilde g_c(r)$ is an increasing function so that there is no shock, and the edge is again given by \eqref{edge0}
(which as we will see below is a general result in the absence of shocks).

The total density $\rho_s$ can be obtained parametrically by taking the derivative of \eqref{r_repulsive1}-\eqref{r_repulsive2} w.r.t. $x$.
One should first solve for $c$ and then plot, for $r \in [-1/2,1/2]$ 
\be \label{rhos_param}
\rho_s = \frac{\mu}{v_0} \frac{\sin(c/2)}{g'_c(r)} \quad , \quad 
x = \frac{v_0}{\mu} \frac{ g_c(r)}{\sin(c/2)} \;,
\ee
for $\kappa<v_0$, and the same expression for $\kappa>v_0$ with $g_c$ replaced by $\tilde g_c$ and the sine function replaced by a $\sinh$.
Similarly one obtains $s(x)$ parametrically from $s=-U(r)/v_0$. This leads to
$\rho_d=-U'(r) \rho_s/v_0$, i.e.
\be \label{rhod_param}
\rho_d = \frac{\mu}{v_0} \frac{\sin(cr)}{g'_c(r)} \quad , \quad 
x = \frac{v_0}{\mu} \frac{ g_c(r)}{\sin(c/2)} \;,
\ee
for $\kappa<v_0$, with the same replacement as above for $\kappa>v_0$.
\\


{\bf Behavior at the edge}.
Let us now see how these densities behave near the edge. Writing $r=\frac{1}{2}-\epsilon$, one has in the case $\kappa<v_0$, $g_c'(\frac{1}{2}-\epsilon) = 2c \cos(c/2) + \epsilon c^2 \sin(c/2)$, which inserting in \eqref{rhos_param} leads to
\be
\rho_s \simeq \frac{\mu}{v_0} \frac{1}{\frac{2c}{\tan(c/2)}+\epsilon c^2} \simeq \frac{\mu}{v_0} \frac{1}{\frac{4\kappa}{v_0}+\epsilon c^2} \simeq \frac{\mu}{4\kappa} \left( 1 - \frac{c^2 v_0}{4\kappa} \epsilon \right)
\ee
(for $\kappa>v_0$ the correction term has a $+$ sign instead). Using that for $r\to 1/2$, $r \simeq \frac{1}{2} - \rho_s(x_e) (x_e-x)$, we obtain, for any $v_0$ and $\kappa$,
\be \label{rho_edge_a1} 
\rho_s(x) 
\simeq \frac{\mu}{4\kappa} \left(1+\frac{C v_0 \mu }{16\kappa^2} (x_e-x)\right) \;.
\ee 

For any $v_0$ and $\kappa$, the density $\rho_s(x)$ has a finite limit at the edge. However, the behavior in the bulk differs in the two cases.
When $\kappa<v_0$, $C$ is negative and thus the density has (local) maxima at the edges of the support. Instead, when $\kappa>v_0$, the edges are (local) minima. Recall that in the non-interacting case, for $a=\gamma/\mu=1/2$, the density has square root divergences at the edges. For $0<\kappa<v_0$, there is no divergence but there is still an accumulation of particle near the edge. However when $\kappa>v_0$, the interaction "hides" the effect of the activity and we do not observe an accumulation of particles at the edges anymore.

Performing a similar expansion for $\rho_d(x)$ and using $\rho_\pm=\frac{\rho_s \pm \rho_d}{2}$ we obtain
\bea \label{rhopm_edge_a1} 
\rho_+(x) &\simeq& \frac{\mu}{4\kappa} \left(1+\big(\frac{C v_0 \mu }{16\kappa^2} - \frac{\mu}{4v_0} \big) (x_e-x)\right) \;, \\
\rho_-(x) &\simeq& \frac{\mu^2}{16 v_0 \kappa} (x_e-x) \;. \nonumber
\eea
Thus there are only $+$ particles at $x_e$ (resp. $-$ particles at $-x_e$), while $\rho_-(x)$ vanishes linearlyat the edge. 

\subsubsection{Attractive case $\kappa=-\bar \kappa<0$} 

We now consider the case where $b=- 2 \bar \kappa<0$. The results of the previous section for $C=- c^2<0$ still hold in this case, i.e. one has
\be \label{r_attractive}
\frac{\mu x}{v_0} \sin(c/2)  = g_c(r) \quad , \quad g_c(r)=\sin (cr) + cr \cos(\frac{c}{2}) \quad , \quad \frac{\tan(c/2)}{c/2}= -\frac{v_0}{\bar\kappa} \;,
\ee 
where now $c$ varies between $\pi$ for $v_0/\bar\kappa\to+\infty$ and $2\pi$ for $v_0/\bar\kappa=0$. We now need to study the function $g_c(r)$ for $r \in [0,1/2]$ and for $c \in [\pi,2 \pi]$. 
It increases as $\sim (1+ \cos(\frac{c}{2})) cr$ at small $r$, up to $r=r_c= \frac{\pi}{c} - \frac{1}{2}$ where it has a maximum of value 
\be 
g_c(r_c)= g_c^* = \sin \left(\frac{c}{2}\right)+\left(\pi - \frac{c}{2}\right) \cos \left(\frac{c}{2}\right) \;.
\ee 
Since $r_c < 1/2$ for $c> \pi$ we see that there is always a shock in the attractive case: in order for $r(x)$ to be monotonous, it should have a discontinuity at the edge such that $r(x_e^-)<1/2$ and $r(x_e^+)=1/2$. This means that $x_e$ is not necessarily equal to $\frac{v_0-\bar\kappa}{\mu}$ (in fact it is always {\it larger}, see below) and the density has delta peaks at $\pm x_e$. Since the shock should occur before (or at) the maximum of the curve $g_c(r)$, we obtain the bounds
\be 
r(x_e^-) \leq r_c \quad , \quad \frac{\mu x_e}{v_0} \sin(c/2)  = g( r(x_e^-)) \leq g_c^* \;.
\ee 

Let us now determine the position of the shock. First of all, note that when $\bar\kappa\geq 2v_0$, the noise is not able to compete with the attraction between particles, and the density is simply a delta peak at $x=0$ (i.e. $x_e=0$ in this case). As explained in \cite{activeRDshort}, in the presence of shocks the equations \eqref{stationary_eq} should be interpreted as
\bea \label{eqdepart_true} 
&& v_0 s'(x)  = \bar \kappa \, [r(x^+)+r(x^-)] r'(x) + V'(x)r'(x) \;,  \\
&& v_0 r'(x) = \bar \kappa [r(x^+)+r(x^-)] s'(x) - 2 \gamma s(x) + V'(x)s'(x) \;.
\eea 
Integrating between $x_e^-$ and $x_e^+$ this leads to,
\bea \label{eq_delta_r}
&& v_0 \Delta s = \bar \kappa (r(x_e^-)+r(x_e^+)) \Delta r + \mu x_e \Delta r \;,  \\
&& v_0 \Delta r = \bar \kappa (r(x_e^-)+r(x_e^+)) \Delta s + \mu x_e \Delta s \;, \nonumber
\eea 
where $\Delta r= r(x_e^+)-r(x_e^-)$ and similarly for $\Delta s$. This implies $\Delta r= \Delta s$ and
\be
v_0 = \bar \kappa (r(x_e^-)+r(x_e^+)) + \mu x_e \;,
\ee
i.e.
\be \label{shock_pos}
\frac{\mu x_e}{v_0} = 1 - \frac{\bar \kappa}{v_0} \big(\frac{1}{2}+r(x_e^-)\big) \;.
\ee
Eq. \eqref{shock_pos} holds for any $a$. Setting $r(x_e^-) \to 1/2$ in this equation one recovers the value of the edge in the
absence of shock, namely $x_e= x_e^0 = \frac{v_0 - \bar \kappa}{\mu}$. Now in presence of
a shock, since 
$r(x_e^-) < 1/2$ one has that $x_e>x_e^0$. Thus because of the presence of the shock the gas can expand 
further. This is a general result valid for any $a$, which is in contrast with the behavior both in the absence of external potential and for a linear potential, where the support was infinite in the absence of shocks. It arises here because the restoring force increases with the distance to the origin. When a cluster forms, the attraction felt by the edge particles is reduced and $x_e$ increases so that this is compensated by the external force.

This result can also be obtained through a physical argument, by assuming that there is a cluster of particles at position $x_e$ containing a fraction $\Delta r = \frac{1}{2}-r(x_e^-)$ of particles, all in the $+$ state and writing the condition for dynamical equilibrium is
\be
v_0 - \mu x_e -\bar \kappa (1-\Delta r) = 0 \;.
\ee

Going back to $a=1/2$ and combining with \eqref{r_attractive} to eliminate $x_e$ and the parameters other than $c$, one obtains
\be
\frac{g_c(r(x_e^-))}{\sin(c/2)} = 1 - \frac{\bar \kappa}{v_0} \big(\frac{1}{2}+r(x_e^-)\big) = 1 + \frac{c}{2 \tan(c/2)} \big(\frac{1}{2}+r(x_e^-)\big) \;,
\ee
and thus $r(x_e^-)$ is determined by 
\be \label{eq_edge_attractive}
h_c(r(x_e^-)) = h_c\big(\frac{1}{2}\big) \quad , \quad h_c(r)=\sin (cr) + \frac{cr}{2} \cos \big(\frac{c}{2}\big) \;,
\ee
where we recall that $c\in[\pi,2\pi]$. The function $h_c(r)$ increases at small $r$ (with $h_c(0)=0$) and has a maximum at $r_{\rm max}<\frac{1}{2}$ for any $c\in[\pi,2\pi]$. Thus, \eqref{eq_edge_attractive} has a solution as long as $h_{c^*}(\frac{1}{2})>0$, i.e. for any $\pi<c<c^*=4.57786...$, where $c^*$ is solution of $h_{c^*}(\frac{1}{2})=0$, i.e.
\be
\frac{\tan (c^*/2)}{c^*/2} = -\frac{1}{2} \;,
\ee
which is consistent with the condition $\bar \kappa<2v_0$ (for $\bar \kappa \geq 2v_0$ the density is a delta peak, as mentioned above). Eq. \eqref{eq_edge_attractive} allows to determine the weight of the delta peaks at $\pm x_e$, and from it, using \eqref{r_attractive}, we obtain
the position of the edge as
\be 
x_e = \frac{v_0}{\mu \sin(c/2)} g_c(r(x_e^-)) \;.
\ee 
Inside the interval $[-x_e,x_e]$, $r(x)$ is still given by \eqref{r_attractive}, and the densities $\rho_s$ and $\rho_d$ are obtained parametrically using \eqref{rhos_param} and \eqref{rhod_param}.
\\

The results of this section are compared to numerical simulations for increasing values of $N$ in Fig.~\ref{fig_a1}. The agreement is very good at large $N$.


\subsection{General case (arbitrary $a$)} \label{sec:general}

Obtaining an explicit expression for $U(r)$ for general values of $a$ turns out to be more challenging. It is however possible to derive some interesting properties of the stationary density without solving the equations explicitly. In this section $\kappa$ is of arbitrary sign.

We start by determining the support of the density $[-x_e,x_e]$, assuming that no shocks are present. In the general case, one has for $r>0$, $U'(r)=-\sqrt{G(U(r))}$ and from the boundary conditions $U(1/2)=0$. From \eqref{G_expression} one sees that $G(0)=v_0^2$ for any parameters, hence $U'(1/2)=- \sqrt{G(0)} = - v_0$ which, using \eqref{U_harmonic}, implies that
\be \label{xe_general}
x_e=\frac{v_0+\kappa}{\mu}
\ee
for any values of the parameters (as long as there are no shocks). Note that $x_e$ vanishes for $\kappa = - v_0$. This coincides with \eqref{shock_pos} when $r(x_e^-)=1/2$, i.e. in the absence of shocks. For $\kappa<-v_0$ the density is a single delta peak at $x=0$.

\subsubsection{Presence of shocks}

Although in general we do not have an explicit relation between $r(x)$ and $x$, it is possible to know whether or not shocks are present in the stationary state. Let us recall the general relation \eqref{eq_xU}
\be \label{eq_xU_shocks}
\mu x(r) = 2 \kappa r - U'(r) \, .
\ee
Shocks will appear if $x(r)$ is a non-monotonous function of $r\in[-1/2,1/2]$. Taking the derivative w.r.t. $r$ leads to
\be
\mu x'(r) = 2 \kappa - U''(r) = 2\kappa - \frac{1}{2} G'(U(r)) \, .
\ee
Thus shocks are present if there exists a value of $u\in[0,u_0]$ such that $G'(u)=4\kappa$. Recall that (for $a\neq 1$)
\be \label{Gprime}
G'(u) = - \frac{1}{a} \left( \frac{v_0^2}{u_0} + \frac{b}{a-1} \right) \left( \frac{u}{u_0} \right)^{\frac{1}{a}-1} + \frac{b}{a-1} \;,
\ee
where we also recall that $a=\gamma/\mu$ and $b=4 a \kappa$.
We have (for $a\neq 1$)
\be
G''(u) = -\frac{1}{a^2 u_0} \left( (1-a) \frac{v_0^2}{u_0} -b \right) \left( \frac{u}{u_0} \right)^{\frac{1}{a}-2} \;.
\ee
$G''(u)$ can be positive or negative depending on the parameters, but it always has a constant sign on $[0,u_0]$, and thus $G'(u)$ is monotonous. Additionally
\be
G'(u_0) = 4\kappa - \frac{\mu v_0^2}{\gamma u_0} < 4\kappa \;.
\ee
Therefore shocks are present if and only if $G'(0)>4\kappa$. Let us start with the repulsive case. One has
\be
G'(0) = \begin{dcases} - \frac{b}{1-a} = - 4\kappa \frac{\gamma}{\mu-\gamma}<0<4\kappa \quad {\rm for} \ a<1 \;, \\
-\infty \hspace{4.4cm} {\rm for} \ a>1 \;. \end{dcases}
\ee
Therefore $G'(u)<4\kappa$ for all $u\in[0,u_0]$, so that there are never any shocks in the repulsive case. 

Let us now turn to the attractive case $\kappa= - \bar \kappa<0$. Shocks are present if and only if $G'(0)>- 4 \bar \kappa$.
For $a<1$,
i.e. $\gamma < \mu$, one has
\be
G'(0) = - \frac{b}{1-a} = 4\bar \kappa \frac{\gamma}{\mu-\gamma}>0>-4\bar \kappa \;,
\ee
so that shocks are always present. The case $a>1$ is less straightforward, since in this case $G'(0)=\pm\infty$ where
the sign depends on the value of the parameters (see \eqref{Gprime}). The condition for shocks to be present is that $G'(0)=+\infty$, i.e.
\be \label{u0star}
u_0>\frac{a-1}{-b} v_0^2 = u_0^* \;.
\ee
Recall that $u_0$ is the solution of
\be
I(u_0)=\frac{1}{2} \quad , \quad I(u_0) = u_0 \int_0^1 \frac{dw}{\sqrt{v_0^2(1-w^{1/a})+\frac{bu_0}{a-1}(w-w^{1/a})}} \;,
\ee
where $I(u_0)$ is an increasing function of $u_0$ with $I(0)=0$. Thus $u_0>u_0^*$ iff $I(u_0^*)<1/2$. One has
\be
I(u_0^*) = \frac{u_0^*}{v_0} \int_0^1 \frac{dw}{\sqrt{1-w}} = 2 \frac{u_0^*}{v_0} = \frac{1}{2} \big(1-\frac{\mu}{\gamma}\big) \frac{v_0}{\bar\kappa} \;,
\ee
and thus the condition for the existence of a shock is
\be \label{ineq_shock}
\frac{\mu}{\gamma}>1-\frac{\bar\kappa}{v_0} \;.
\ee
Hence in the attractive case case there is a region without shocks (left part of region I in Fig.~\ref{phase_diagram}) and regions with shocks (III and IV). In the absence of shocks, the position of the edge is given by \eqref{xe_general}. The analysis of the edge behaviour will be performed below, along with the repulsive case. If shocks are present, the equations \eqref{eqU_G}-\eqref{eq_xU} are still valid and determine $r(x)$ for $|x| < x_e$. The position of the edge is then determined by Eq.~\eqref{shock_pos}, which is valid for any value of $a$. Combining with \eqref{eqU_G}-\eqref{eq_xU} we obtain that 
$0< r(x_e^-) < \frac{1}{2}$ is solution of the equation
\be \label{rxe} 
\bar \kappa (\frac{1}{2} - r(x_e^-)) - U'(r(x_e^-)) = v_0 \;.
\ee 
Once this solution is obtained $x_e$ is determined from \eqref{shock_pos}. 
Note that $r(x_e^-)=1/2$ is always solution of \eqref{rxe}, however
when there is another solution $0< r(x_e^-) < \frac{1}{2}$ the latter is the
correct solution. 
Since $U'(r)<0$ for $r>0$ the equation \eqref{rxe} is valid only for $\bar \kappa<2 v_0$.
The case $\bar \kappa \geq 2 v_0$ corresponds to the phase where the density is
a single delta peak at $x=0$ (with $r(x_e^-) \to 0^+$ as $\bar \kappa \to 2 v_0^-$).
The limit case $r(x_e^-) \to \frac{1}{2}^-$ corresponds to the disappearance of the shocks.

We have analyzed these equations above for $a=1/2$
and we will not study here their general solution. 
As was discussed in the case $a=1/2$, we find that $x_e$ is larger in the presence of shocks, due to the convexity of the confining potential.
\\

\subsubsection{Sign of $C$ and the special line $\frac{\mu}{\gamma}=1+ \frac{\kappa}{v_0}$ } 

As we have already seen in the case $a=1/2$, and as we will discuss more generally below, the sign of $C$ influences the slope of the density near the edges of the support when there are no shocks. 

\begin{figure}
    \centering
    \includegraphics[width=0.35\linewidth, trim={3cm 1cm 2.5cm 0.8cm},clip]{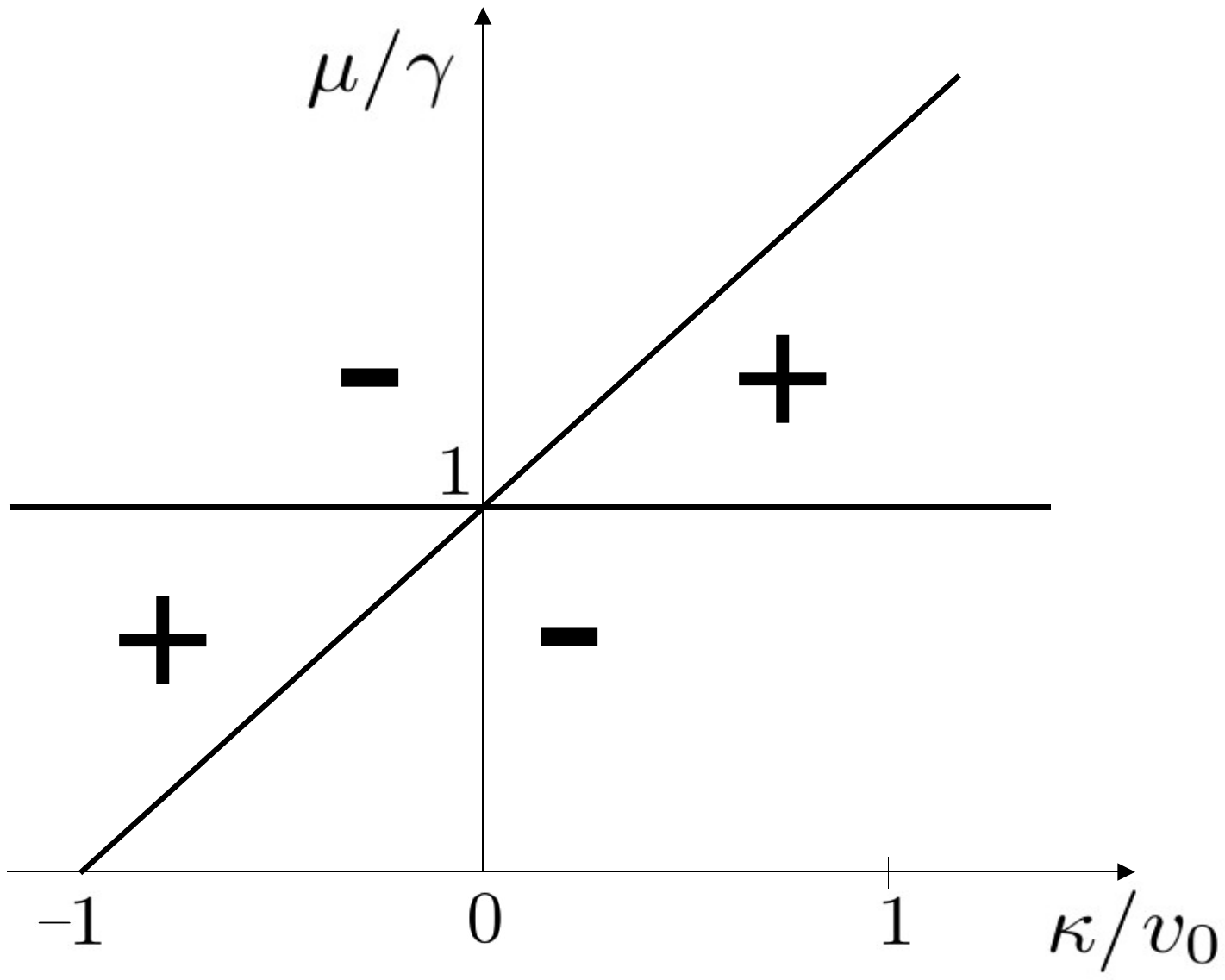}
    \caption{Sign of the integration constant $C$. It vanishes on the line $\frac{\mu}{\gamma}=1 + \frac{\kappa}{v_0}$.}
    \label{signofC}
\end{figure}

The sign of $C$ can be obtained from the above considerations. 
Indeed, for $a\neq 1$, one has, rewriting \eqref{C_expression} using $u_0^*=\frac{1-a}{b}v_0^2$ ,
\be
C = v_0^2 u_0^{-1/a} \big( \frac{u_0}{u_0^*}-1 \big) \;.
\ee
There are 2 cases to distinguish. If $b>0$ and $a>1$, or $b<0$ and $a<1$, then $u_0^*<0$, and since $u_0>0$ one has $C<0$. In the opposite case, one has ${\rm sgn} (C) = {\rm sgn} \, (u_0-u_0^*)$, and the computation above shows that $C>0$ iff the inequality $I(u_0^*)<1/2$ is satisfied, which is equivalent to \eqref{ineq_shock} for $\kappa=-\bar \kappa<0$ and to $\frac{\mu}{\gamma}<1+\frac{\kappa}{v_0}$ for $\kappa>0$. This is summarized in
Fig.~\ref{signofC}. 

It is possible to determine exactly the shape of the density on the line in the phase diagram where $C$ vanishes.
Let us set $C=0$ in \eqref{G_expression}, leading to 
\be
G(u) = \frac{b}{a-1}u + v_0^2 \;.
\ee
Then \eqref{eqU_G} becomes
\be
\frac{1}{2} - r = \int_{0}^{U(r)} \frac{du}{\sqrt{\frac{b}{a-1}u + v_0^2}} = \frac{2(a-1)}{b} \left(\sqrt{\frac{b}{a-1}U(r)+v_0^2}-v_0 \right) \;,
\ee
and thus
\be
U(r) = \frac{b}{4(a-1)} (\frac{1}{2}-r)^2 +v_0(\frac{1}{2}-r) \;.
\ee
The condition $U'(0)=0$ (see \eqref{signUprime}) then implies
\be \label{signC}
\frac{b}{4(1-a)} = v_0 \quad \text{i.e.} \quad \frac{\mu}{\gamma} = 1 + \frac{\kappa}{v_0} \;.
\ee
This condition defines a line in the phase diagram for which $C=0$ and which coincide with the one
obtained above, see Fig. \ref{signofC}. One can determine the densities on this line. Inserting into \eqref{U_harmonic}, this leads to a uniform total
density within the support
\be 
r(x) = \frac{\mu x}{2 (\kappa + v_0)} \quad , \quad \rho_s(x)= \frac{\mu}{2 (\kappa + v_0)}  \quad , \quad \rho_d(x)=  \frac{\mu^2 x}{2(\kappa+v_0)^2} 
\quad , \quad x \in [- x_e,x_e] 
\quad , \quad x_e = \frac{v_0 + \kappa}{\mu} \;. \label{edge0}
\ee 
where $x_e$ is given by \eqref{xe_general}. These results coincide with \eqref{r_square_ahalf} and \eqref{s_square_ahalf} for $a=1/2$ and $\kappa=v_0$. Note that there are never shocks along this line.

Finally, $C$ also changes sign on the line $a=1$, which is more complicated to analyze. We will not consider it here. 

\subsubsection{Edge behaviour (in the absence of shocks)}

We now turn to the edge properties of the density. In this section the sign of $\kappa$ is still arbitrary, but we restrict to the case where there is no shock.



The behaviour of the density near the edges can be obtained by taking the derivative of \eqref{xparamU} w.r.t $x$. This leads to
\be \label{eq_rho_U} 
\rho_s(x) = \frac{2 \gamma U}{v_0^2 - U'^2}= \frac{2 \gamma U}{v_0^2 - G(U)}
= \begin{dcases} \frac{2 \gamma}{\frac{b}{1-a} - C U^{\frac{1}{a}-1} } \quad \text{for } a \neq 1 \;, \\
-\frac{2 \gamma}{b\ln U + C} \quad \quad \text{for } a = 1 \;. \end{dcases}
\ee
Since $U(1/2)=0$, there are two cases: for $a\geq 1$, one has $\rho_s(x) \to 0$ as $x\to x_e$ and the density vanishes at the edge, while for $a<1$ the density has a finite limit at the edge
\be
\rho_s(x_e) = \frac{2 \gamma}{b} (1 - a) = \frac{\mu - \gamma}{2 \kappa} \;.
\ee 
More precisely, using $U(r)\simeq v_0(\frac{1}{2}-r) \simeq v_0 \rho_s(x_e) (x_e-x)$ near $r=1/2$, \eqref{eq_rho_U} leads to, for $a<1$, i.e. $\mu/\gamma>1$,
\be \label{rho_edge_C}
\rho_s(x) \simeq \frac{\mu - \gamma}{2 \kappa} \left(1 + C \frac{\mu-\gamma}{4 \gamma\kappa} v_0^{\frac{\mu}{\gamma}-1} (\frac{1}{2}-r)^{\frac{\mu}{\gamma}-1}\right)
\simeq \frac{\mu - \gamma}{2 \kappa} \left(1 + \frac{C}{2\gamma} \left(\frac{\mu-\gamma}{2\kappa}\right)^{\mu/\gamma} v_0^{\frac{\mu}{\gamma}-1} (x_e-x)^{\frac{\mu}{\gamma}-1}\right) \;.
\ee 

The sign of $C$ was discussed in the previous subsection, see Fig. \ref{signofC}, where we found that whenever $\mu/\gamma<1+\kappa/v_0$, one has $C>0$, which according to \eqref{rho_edge_C} means that the density has a local minimum at the edge. If instead $\mu/\gamma>1+\kappa/v_0$, then $C<0$, leading to a local maximum at the edge. Note however that this is only valid for $a<1$, i.e. $\mu/\gamma>1$, and in the absence of shocks, i.e. in the repulsive case $\kappa>0$. This change of edge behaviour marks the difference between the regimes IIa and IIb in Fig.~\ref{phase_diagram}.
\\


Let us now determine how $\rho_s(x)$ vanishes at the edge for $a>1$. Indeed from \eqref{eq_rho_U} we find that for $x$ near $x_e$,
\be \label{asympt0}
\rho_s(x) \simeq \frac{2 \gamma}{|C|}  U^{1-\frac{1}{a}} \;,
\ee 
where we have used that $C<0$ in this regime. Let us assume that 
\be 
\rho_s(x) \simeq A (x_e-x)^\alpha \;.
\ee 
Since $U(1/2)=0$ and $U'(1/2)=-v_0$, see above, this implies 
\be \label{Uedge_x}
U(r) \simeq v_0\big(\frac{1}{2}-r\big) = v_0 \frac{A}{\alpha+1}  (x_e-x)^{\alpha+1} \;.
\ee 
Plugging into \eqref{asympt0} gives
\be 
A (x_e-x)^{\alpha} \simeq \frac{2 \gamma}{|C|} \big(v_0 \frac{A}{\alpha+1}\big)^{1-\frac{1}{a}} (x_e-x)^{(\alpha+1)(1- \frac{1}{a})} \;.
\ee 
This implies $\alpha=a-1$, and the behavior near the edge, for $a>1$, i.e. for $\gamma>\mu$, is given by
\be \label{rhos_edge}
\rho_s(x) \simeq \left(\frac{2\gamma}{|C|}\right)^a \left(\frac{v_0}{a}\right)^{a-1} (x_e-x)^{a-1} = \gamma \left(\frac{2}{|C|}\right)^{\frac{\gamma}{\mu}} (\mu v_0)^{\frac{\gamma}{\mu}-1} (x_e-x)^{\frac{\gamma}{\mu}-1} \;.
\ee 
This is the same exponent as in the non-interacting case $\kappa=0$ (with a different amplitude which depends on $\kappa$). 
This is in contrast with the case $\gamma<\mu$, for which the density for $\kappa=0$ diverges at the edges, while it has a finite limit for $\kappa>0$.
\\

Finally, in the marginal case $a=1$, the density again vanishes at the edge but not with an algebraic exponent. One has in this case
\be
\rho_s(x) \simeq -\frac{2\gamma}{b\ln U} \simeq -\frac{2\gamma}{b\ln (\frac{1}{2}-r)} \;.
\ee
We restrict ourselves to the repulsive case $b>0$, since in the attractive case there is always a shock for $a=1$. Using that $\rho_s(x)=r'(x)$ and integrating we obtain
\be
-(\frac{1}{2}-r) \ln (\frac{1}{2}-r) \simeq \frac{2\gamma}{b} (x_e-x) \;,
\ee
which implies
\be
\frac{1}{2}-r(x) \simeq -\frac{2\gamma}{b} \frac{(x_e-x)}{\ln (x_e-x)} = -\frac{\mu}{2\kappa} \frac{(x_e-x)}{\ln (x_e-x)} \;.
\ee
Taking the derivative, we obtain at leading order the density near the edge
\be \label{logbehavior} 
\rho_s(x) \simeq \frac{\mu}{2\kappa} \frac{1}{\big(\ln (x_e-x)\big)^2} \;.
\ee
It is worth noting that this inverse logarithmic divergence only occurs for $\kappa >0$ (in the non-interacting
case the density is uniform on $x \in [-v_0/\mu,v_0/\mu]$ for $a=1$, i.e. $\gamma=\mu$).
\\

{\bf Edge behaviour of $\rho_\pm(x)$.} Finally, let us briefly consider the densities of $+$ and $-$ particles independently. For $x$ near $x_e$, one has
\be 
\rho_d(x) = - \frac{1}{v_0} \partial_x U(r(x)) \simeq - \frac{1}{v_0} \partial_x [v_0(\frac{1}{2}-r)] = \rho_s(x) \;.
\ee 
Hence $\rho_+(x)\simeq \rho_s(x)$, while $\rho_-(x)$ vanishes at the edge with a larger exponent. In the case $\kappa=0$, $\rho_-(x)$ vanishes with an exponent $\frac{\gamma}{\mu}$ near $x_e$. In general we have
\be \label{rho_-0}
\rho_-(x) = \frac{1}{2}(\rho_s(x)-\rho_d(x)) = \frac{1}{2}(1+\frac{U'(r)}{v_0})\rho_s(x) \;.
\ee
For $a<1$, one has from \eqref{Usecond} $U''(1/2)=\frac{1}{2}G'(0)=\frac{b}{2(a-1)}$, and thus near $x_e$,
\be
U'(r) \simeq -v_0 +\frac{b}{2(1-a)}(\frac{1}{2}-r) \simeq -v_0 +\frac{b}{2(1-a)} \frac{\mu-\gamma}{2\kappa} (x_e-x) \simeq -v_0 + \gamma (x_e-x) \;.
\ee
Inserting in \eqref{rho_-0} we obtain, for $\gamma<\mu$,
\be \label{rho_-1}
\rho_-(x) \simeq \frac{1}{2} \frac{\gamma}{v_0} (x_e-x) \rho_s(x) \simeq \frac{\gamma(\mu-\gamma)}{4\kappa v_0} (x_e-x) \;.
\ee

In the case $a>1$, i.e. $\gamma>\mu$, one has instead
\be
U'(r) = -\sqrt{G(U(r))} \simeq - v_0 (1+\frac{C}{2v_0^2}U(r)^{1/a}) \;.
\ee
Combining with \eqref{Uedge_x}, \eqref{rhos_edge} and \eqref{rho_-0} one obtains,
\be
\rho_-(x) \simeq \frac{|C|}{4v_0^2}U(r)^{1/a} \rho_s(x) \simeq \frac{\mu}{2v_0} (x_e-x) \rho_s(x) \simeq \frac{\gamma}{2} \left(\frac{2\mu}{|C|}\right)^{\frac{\gamma}{\mu}} v_0^{\frac{\gamma}{\mu}-2} (x_e-x)^{\gamma/\mu}\;.
\ee
Similarly for $a=1$, i.e. $\gamma=\mu$, one obtains
\be
\rho_-(x) \simeq -\frac{b}{4v_0^2} U(r) \ln U(r) \rho_s(x) \simeq \frac{\mu^2}{4v_0\kappa} \frac{x_e-x}{\big(\ln (x_e-x)\big)^2} \;,
\ee 
where once again (as in \eqref{logbehavior}) the logarithmic behaviour is a special feature of the interacting case $\kappa>0$ (since it vanishes linearly when $\kappa=0$).

For $\gamma>\mu$, the exponent is again the same as in the $\kappa=0$ case. In general, we observe a difference of 1 in the exponent between $\rho_+(x)$ and $\rho_-(x)$, as in the non-interacting \cite{DKM19} and the active DBM cases \cite{TouzoDBM2023}.
\\


\subsection{Phase diagram}

The results of this section are summarized in Fig.~\ref{phase_diagram}, which shows a diagram of the different regimes for the stationary density. 
These results are confirmed by numerical simulations, where we solve numerically the stochastic equation of motion for $N$ RTP's (with $N$ up to $10^3$) and
measure the stationary densities by averaging over a large time window. We find that 
there are 5 phases in total:
\begin{itemize}
    \item Phase I: A smooth 
    phase which exists both in the repulsive ($\mu/\gamma<1$) and in the attractive case ($\mu/\gamma<1-\bar\kappa/v_0$), in which the density vanishes at the edges with an exponent $\frac{\gamma}{\mu}-1$ identical to the exponent of the non-interacting case.
    \item Phase IIa: A strongly repulsive phase ($1<\mu/\gamma<1+\kappa/v_0$), in which the density has a discontinuity at the edges, but is minimal at the edges.
    \item Phase IIb: A strongly active phase ($\mu/\gamma>1+\kappa/v_0$), in which the density has a discontinuity at the edges and is maximal at the edges.
    \item Phase III: A strongly attractive phase ($\mu/\gamma>1-\bar\kappa/v_0$ and $\bar\kappa/v_0<2$), in which the density has delta peaks at the edges.
    We find in our numerical simulations with $N=1000$ that the density is maximal at the edges (independently of the sign of $C$).
    \item Phase IV: A condensed phase ($\bar\kappa/v_0>2$), in which the density is a delta peak at $x=0$.
\end{itemize}

The passage from phase IIa to IIb is only a crossover since it also occurs at finite $N$ (see Fig.~\ref{phase_diagram}), while
the other phases have distinct features which only occur for $N\to+\infty$. At finite $N$, we find from our numerical simulations that both the jumps at the edge of the density in phases IIa and IIb and the shocks in phase III have a small width which decays to zero as $N$ increases, see Fig \ref{fig_a1}. 
In the case of the shocks, this broadening is due to fluctuations in the number of particles contained in the clusters of particles located at the edge, which lead to fluctuations in their position.
The phase IV is special: if at finite $N$ we measure the density in the reference frame of
the center of mass, it will always be a delta peak for $\kappa \leq -2 v_0$ for any $N\geq 2$. 
The absolute total density however remains smooth at finite $N$ since the center of mass reaches
a non trivial stationary distribution.


\subsection{Limit of small $\gamma$ and fixed points}

\begin{figure}
    \centering
    \includegraphics[width=0.32\linewidth]{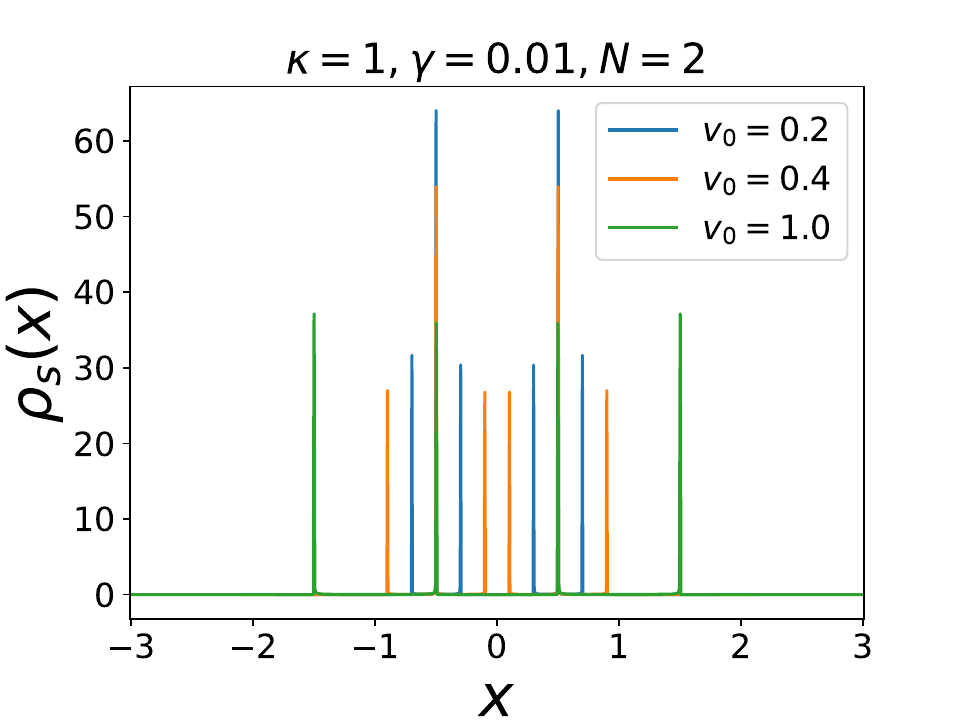}
    \includegraphics[width=0.32\linewidth]{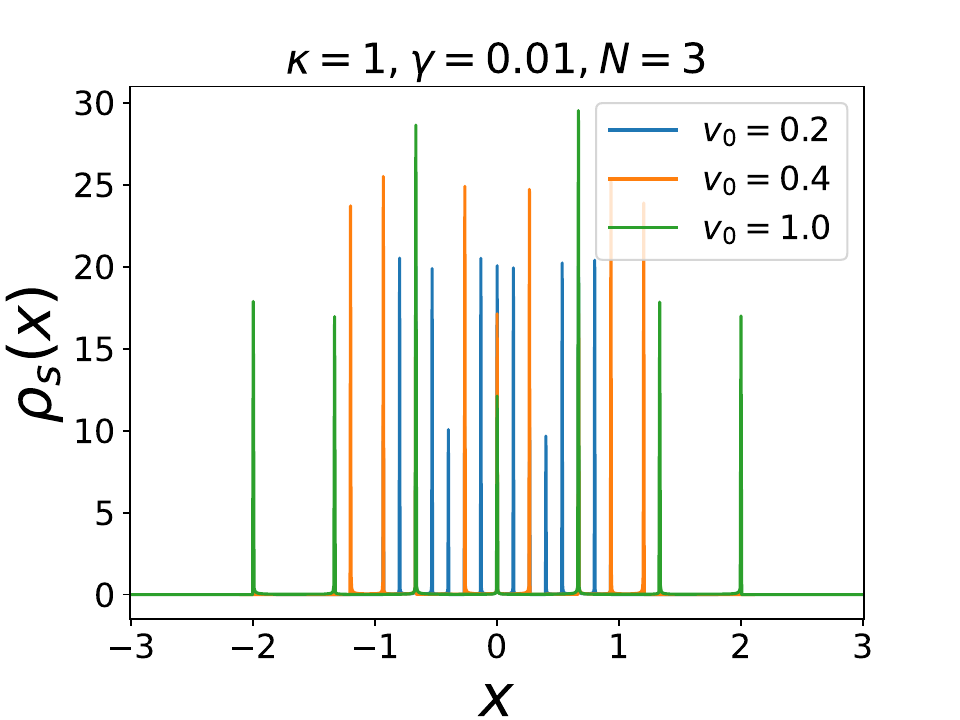}
    \includegraphics[width=0.32\linewidth]{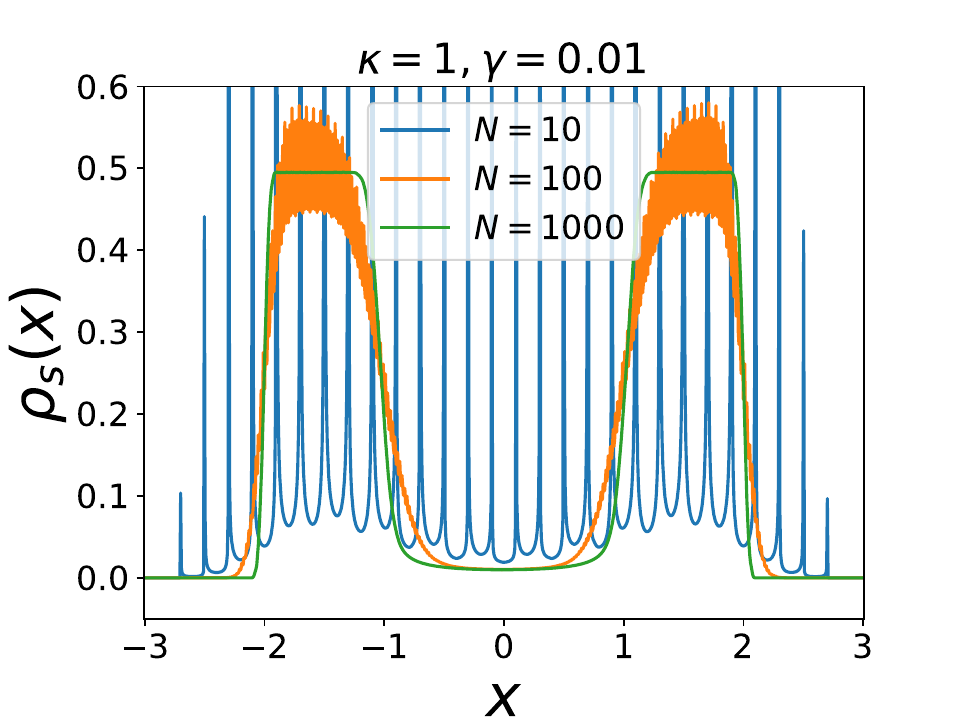}
    \caption{Plots of the stationary density $\rho_s(x)$ in the repulsive case with $\kappa=1$, $\mu=1$ and $\gamma=0.01$. Left and central panels: plots for $N=2$ and $3$ with different values of $v_0$. The density is mostly composed of delta peaks corresponding to the different fixed points of the system. The delta peaks closest to the center disappear for $v_0>\kappa/N$ (e.g. we go from 6 to 4 peaks for $N=2$). Right: Plots for $v_0=1$ and increasing values of $N$. As $N\to+\infty$, $\rho_s(x)$ converges to a smooth density on $[-\frac{v_0+\kappa}{\mu},\frac{v_0+\kappa}{\mu}]$. In the limit $\gamma\to 0$, this density is uniform on $[-\frac{v_0+\kappa}{\mu},-\frac{v_0}{\mu}]\cup[\frac{v_0}{\mu},\frac{v_0+\kappa}{\mu}]$ and zero on $[-\frac{v_0}{\mu},\frac{v_0}{\mu}]$.}
    \label{figFP}
\end{figure}

For $\gamma \ll \mu$, the stationary density is closely related to the fixed points of the dynamics at fixed $\sigma_i$'s, given by the roots of the set of equations 
\be
V'(x_i) = \mu x_i = \frac{\kappa}{N} \sum_{j=1}^N {\rm sgn} (x_i-x_j) + v_0\sigma_i \quad , \quad i=1,\dots,N \;.
\ee
In the limit $\gamma/\mu \to 0$ the system spends most of its time near these fixed points before the $\sigma_i(t)$
switch to another value. In the stationary state each set of $\sigma_i$ is equiprobable.
At finite $N$, since these fixed points are reached in finite time, they appear as sharp peaks in the density (which become delta peaks for $\gamma=0^+$), see Figure \ref{figFP}.
A similar discussion was detailed in the case of the active DBM in \cite{TouzoDBM2023}. 

The attractive case is quite straightforward to analyze: in the limit $\gamma\to 0^+$, the particles will form two clusters at positions $\pm x_e$ containing respectively all the $+$ and $-$ particles. Each clusters thus contains $N/2$ particles. The position $x_e$ is simply given by the balance of forces
\be
\mu x_e = v_0-\frac{\bar \kappa}{2} \;.
\ee
which is larger than the position of the edge in the absence of shocks $\mu x_e = v_0 -\bar \kappa$, see 
\eqref{xe_general}. For $\bar \kappa\geq 2v_0$, the two clusters merge and $x_e=0$.

The repulsive case is more involved. 
For $v_0=0$, denoting $x_i^0=\frac{\kappa}{2 N \mu} (N+1-2 i)$, there are $N!$ fixed points, which take the form $x_i=x_{\tau(i)}^0$ where
$\tau$ is any permutation of the particles. The particles are thus 
regularly spaced with separation $\frac{2 \kappa}{N \mu}$. For $v_0 < \frac{\kappa}{N}$ each
fixed point is simply shifted as $x_i=x_{\tau(i)}^0 + \frac{v_0}{\mu} \sigma_i$. As $v_0$ increases beyond $\frac{\kappa}{N}$, a natural conjecture is that
there is a transition to a set of $N_+! N_-!$ fixed points where all the $+$ particles are to the right of all the $-$ particles
(which are identical up to permutations in each group). The positions are still given by $x_i=x_{\tau(i)}^0 + \frac{v_0}{\mu} \sigma_i$
but now the permutation $\tau \in S_{N_+} \times S_{N_-} $ only permutes member of each group.

In the limit $N\to +\infty$, the fraction of $+$ and $-$ particles are the same and thus all $+$ particles are evenly spaced on $[\frac{v_0}{\mu},\frac{v_0+\kappa}{\mu}]$ and all $-$ particles are evenly spaced on $[-\frac{v_0+\kappa}{\mu},-\frac{v_0}{\mu}]$. Thus the density in the double limit $N\to+\infty$, $\gamma\to 0$, is
\be \label{rhos_gamma0}
\rho_s(x) = \frac{\mu}{2\kappa} \quad {\rm for} \ \frac{v_0}{\mu}<|x|<\frac{v_0+\kappa}{\mu} \;.
\ee
Note that this regime $\gamma/\mu \ll 1$ is a limit of the phase IIb in Fig. \ref{phase_diagram}.

This result can also be derived by solving \eqref{stationary_eq} for $\gamma=0$, and generalized to a larger class of potentials $V(x)$.
Indeed the first equation in \eqref{stationary_eq} leads to $s'(x)=\frac{V'(x)-2\kappa r(x)}{v_0}r'(x)$. Inserting in the second equation gives
\be
v_0^2 r'(x) = (V'(x)-2\kappa r(x))^2 r'(x) \;,
\ee
i.e. $r'(x)=s'(x)=0$ or
\be
r(x) = \frac{V'(x)\pm v_0}{2\kappa} \quad , \quad s'(x)=\mp r'(x) \;.
\ee
The second equation means that the $+$ and $-$ particles are separated. Let us assume for simplicity $V(x)$ to be convex, twice differentiable
with $V'(\pm \infty)=\pm \infty$. In that case one can define $x_0^\pm$ and $x_e^\pm$ as the roots of
\be 
V'(x_0^\pm)= \pm v_0   \quad , \quad V'(x_e^\pm)= \pm ( v_0 + \kappa) \;.
\ee 
The support of the density has two components, with
\be 
r(x) = \begin{cases}  \frac{V'(x)- v_0}{2\kappa}  \quad , \quad x \in [x_0^+,x_e^+] \;, \\
\frac{V'(x)+ v_0}{2\kappa}  \quad , \quad x \in [x_e^-,x_0^-] \;,
\end{cases} 
\ee 
where $\rho_s(x)=\frac{V''(x)}{2 \kappa}$ in each component, and $\rho_d(x)=\rho_s(x)$ for $x \in [x_0^+,x_e^+] $
and $\rho_d(x)=-\rho_s(x)$ for $x \in [x_e^-,x_0^-] $. Taking $V(x)=\frac{\mu}{2}x^2$, this is compatible with \eqref{rhos_gamma0}.

We have tested these predictions in a numerical simulation of the stationary density for various values of $N$,
for $\gamma \ll \mu$,
see Fig. \ref{figFP}. For low values of $N$ the density is well approximated by a sum of delta peaks corresponding to the fixed points.
One sees that some of these peaks disappear as $v_0$ is increased, as predicted. For large values
of $N$ the density become smooth. As $\gamma \to 0$ a gap region develops where the density vanishes, 
with two uniform square densities on each side, for each $\pm$ species respectively. 

\section{Non-reciprocal active rank diffusions in a linear potential} \label{sec:nonreciprocal}

\subsection{Equation for the rank fields for a general non-reciprocal interaction}

We now consider a more general set of stochastic equation of motion, where the interaction potential between particles $i$ and $j$ still
depends on their rank, but also on their states $\sigma_i(t)$ and $\sigma_j(t)$, with interaction strength 
$\kappa_{\sigma_i(t),\sigma_j(t)}$, where the 
$\kappa_{\pm,\pm}$ are a given set of four parameters
\bea \label{langevingeneral}
\frac{dx_i}{dt} =  \sum_{j=1}^N \frac{\kappa_{\sigma_i(t),\sigma_j(t)}}{N} {\rm sgn}(x_i-x_j) - V_{\sigma_i(t)}'(x_i) + v_0 \sigma_i(t) + \sqrt{2 T} \xi_i(t) \;.
\eea 
Here $\kappa_{\sigma,\sigma'}$ parameterizes the force exerted by a particle in state $\sigma'$ on a particle in state $\sigma$. 
We do not assume here that the matrix $\kappa_{\sigma,\sigma'}$ is symmetric. 

In this case the Dean-Kawasaki equations become
\bea \label{eqrho1general}
\partial_t \rho_\sigma(x,t)  &=&  T \partial_x^2 \rho_\sigma(x,t)  +  \partial_x [\rho_\sigma(x,t)  ( - v_0 \sigma +  V_\sigma'(x) -
\sum_{\sigma'} \kappa_{\sigma, \sigma'} \int dy \rho_{\sigma'}(y,t)  {\rm sgn}(x-y) )] \nonumber \\
&& + \gamma \rho_{- \sigma}(x,t) - \gamma \rho_{\sigma}(x,t) +O\left( \frac{1}{\sqrt{N}} \right) \;.
\eea

We define
\be
h_\pm(x) = \frac{1}{2} \left(r(x) \pm s(x) \right) = \int^x_{-\infty} dy \, \rho_\pm(y,t)\, -\, \frac{1}{4} \;,
\ee
where $r(x)$ and $s(x)$ are defined as in \eqref{rankdef1}, such that
\be 
\rho_{\pm}= \frac{1}{2} (\rho_s \pm \rho_d) = \partial_x \frac{r \pm s}{2} = \partial_x h_\pm \;,
\ee 
and we rewrite the interaction term using
\bea
\int dy \rho_\pm(y,t) {\rm sgn}(x-y) &=& \int dy \partial_y (\frac{r(y,t) \pm s(y,t)}{2} ) {\rm sgn}(x-y)
\\
&=& r(x,t) \pm s(x,t) + [(\frac{r(y,t) \pm s(y,t)}{2} )  {\rm sgn}(x-y)]^{+\infty}_{-\infty} \nonumber \\
&=& r(x,t) \pm s(x,t) 
- \frac{1}{2} (r(-\infty,t) + r(+\infty,t)) - \frac{1}{2} (s(-\infty,t) + s(+\infty,t)) \nonumber \\
&=&  r(x,t) \pm s(x,t) - \frac{1}{2}  \int_{-\infty}^{+\infty} dy \rho_d(y,t) \nonumber \\
&=& 
r(x,t) \pm s(x,t) - \frac{1}{2}  (p_+(t) - p_-(t)) \;. \nonumber
\eea 
For simplicity for now we will assume that we start from $p_+(t=0)=p_-(t=0)=1/2$ and neglect again the noise
at large $N$ so that $p_+(t)=p_-(t)=1/2$ at all times. Hence we obtain at large $N$
\be \label{eqrho2general}
\partial_t \rho_\sigma(x,t) =  T \partial_x^2 \rho_\sigma(x,t)  +  \partial_x [\rho_\sigma(x,t)  ( - v_0 \sigma +  V_\sigma'(x) -
2  \sum_{\sigma'} \kappa_{\sigma, \sigma'} h_{\sigma'}(x,t)  )] + \gamma \rho_{- \sigma}(x,t) - \gamma \rho_{\sigma}(x,t) \;,
\ee
where because of our assumptions ($p_+(t)=p_-(t)=1/2$) one has the boundary conditions $h_\sigma(\pm \infty,t)= \pm 1/4$. We can thus integrate
the equation and obtain
\be \label{eqrho3general}
\partial_t h_\sigma(x,t) = T \partial_x^2 h_\sigma(x,t)  +  ( - v_0 \sigma +  V_\sigma'(x) -
2  \sum_{\sigma'} \kappa_{\sigma, \sigma'} h_{\sigma'}(x,t)  )] \partial_x h_\sigma(x,t) + \gamma h_{- \sigma}(x,t) - \gamma h_{\sigma}(x,t) \;.
\ee

\subsection{A simple example of non-reciprocal interaction}

Let us consider a non-reciprocal interaction of the form
\be 
\kappa_{+-} = - \kappa_{-+} = -b \quad , \quad \kappa_{++} = \kappa_{--} = 0 \;.
\ee 
This means that, if $b>0$, $+$ particles are attracted to $-$ particles, while $-$ particles are repulsed by $+$ particles, both with a force of norm $b/N$ independent of the distance. Note that there is a symmetry ($b\to-b$, $x\to-x$, $h_+\leftrightarrow h_-$). We thus restrict our study to $b>0$, since the case $b<0$ can be obtained by symmetry. 
We look for the stationary measure at $T=0$, in presence of a linear external potential $V'(x)=a \, {\rm sgn}(x)$ (with $a>0$, since in the absence of external potential there is no stationary state and the particles escape to infinity). The stationary equations read
\bea
&& 0 =   ( - v_0  +  a \, {\rm sgn}(x)  +
2 b h_{-} ) h'_+  + \gamma h_{-} - \gamma h_{+} \;, \\
&& 0 =   (  v_0  +  a \, {\rm sgn}(x)  
- 2 b h_{+} ) h'_-  + \gamma h_{+} - \gamma h_{-} \;.
\eea 
This leads to, taking the sum and difference,
\bea 
&& 0 = - v_0 s' + a \, {\rm sgn}(x) r' + 2 b ( h_{-} h'_+ - h_{+} h'_- ) \;, \\
&& 0 = - v_0 r' + a \, {\rm sgn}(x) s' + 2 b ( h_{+} h_- )' - 2 \gamma s \;,
\eea
which can be rewritten as
\bea 
&& 0 = - v_0 s' + a \, {\rm sgn}(x) r' + b r s' - b s r' \;, \label{eqnr1} \\
&& 0 = - v_0 r' + a \, {\rm sgn}(x) s' + b r r' - b s s' - 2 \gamma s  \;. \label{eqnr2}
\eea

The first equation can be rewritten as
\be 
(v_0- b r) \frac{ds}{dr} = (\pm a  - b s ) \;,
\ee 
where $\pm a = a \, {\rm sgn}(x)$. It integrates into
\be 
s= \pm \frac{a}{b} + (b r - v_0) c_\pm \;.
\ee 
We assume for now that there are no shocks outside of $x=0$. In this case, the constant $c_\pm$ is determined by the constraints $s(\pm \infty)=0$ and $r(\pm \infty)=\pm 1/2$. One finds 
\be 
c_+ = \frac{2 a}{b(2 v_0-b)} \quad , \quad c_- = - \frac{2 a}{b(2 v_0+b)} \;,
\ee 
which leads to
\be \label{rel_sr}
s = \begin{dcases} a \frac{1 - 2 r}{b-2 v_0} \quad \hspace{0.23cm} \text{for } x>0 \;, \\
 - a \frac{1+ 2 r}{b+ 2 v_0} \quad \text{for } x<0 \;. \end{dcases}
\ee 
\\

Going back to Eq.~\eqref{eqnr2} and inserting \eqref{rel_sr}, we find, for $x>0$,
\be 
\frac{  \left((2 v_0-b)^2- 4 a^2\right) 
   \left(v_0-b r(x)\right)}{\left(2 v_0-b\right)^2} r'(x) = \frac{2 a \gamma 
   (1-2 r(x))}{2 v_0 - b} \;,
\ee 
that is
\be
\frac{v_0-b r}{1- 2 r} dr = \frac{2 a \gamma (2 v_0- b)}{(2 v_0-b)^2- 4 a^2 }  dx \;,
\ee 
which integrates into
\be \label{eqnr_rx_pos}
b (1- 2 r) + (2 v_0-b) \log(1-2 r)  = - \frac{8 a \gamma (2 v_0- b)}{(2 v_0-b)^2- 4 a^2 } x + C_+ \;,
\ee 
where $C_+$ is a constant that remains to be determined.

The solution, assuming that it exists, decays exponentially at large $x$ and can be written as
\be
1 - 2 r(x) = \frac{2 v_0-b}{b} W\left(\frac{b}{2 v_0-b} e^{-A_+ x +\frac{C_+}{2 v_0 - b}}\right) \quad , \quad A_+=\frac{8 a \gamma}{(2 v_0-b)^2- 4 a^2 } \;,
 \label{Wpositive0} 
\ee
where $W$ is the main branch of the Lambert function, 
which is the real (and largest) root of 
\be 
W(z) e^{W(z)} = z \;,
\ee 
for $z \in [-1/e,+\infty[$, with $W(z) \simeq z$ for $z$ near $z=0$. Thus the density decays as $\rho_s(x)\sim e^{-A_+x}$ for $x\to+\infty$.
\\

For $x<0$ one finds by a similar calculation
\be 
\frac{  \left((2 v_0+b)^2- 4 a^2\right) 
   \left(v_0-b r(x)\right)}{\left(2 v_0+b\right)^2} r'(x) = \frac{2 a \gamma 
   (1+2 r(x))}{2 v_0 + b} \;,
\ee 
that is
\be
\frac{v_0-b r}{1+2r} dr = \frac{2 a \gamma (2 v_0 + b)}{(2 v_0+b)^2- 4 a^2 }  dx \;,
\ee 
which integrates into
\be \label{eqnr_rx_neg}
b (1+2r) - (2 v_0+b) \log(1+2 r)  = - \frac{8 a \gamma (2 v_0+b)}{(2 v_0+b)^2- 4 a^2 } x + C_- \;,
\ee 
where $C_-$ is another constant to be determined.

Assuming again that a solution exists, it reads
\be \label{Wnegative0} 
1 + 2 r(x) = -\frac{2 v_0+b}{b} W\left(-\frac{b}{2 v_0+b} e^{A_- x -\frac{C_-}{2 v_0 + b}}\right) \quad , \quad A_-=\frac{8 a \gamma}{(2 v_0+b)^2- 4 a^2} \;.
\ee
Thus the density decays as $\rho_s(x)\sim e^{A_-x}$ for $x\to-\infty$.

We see that $A_+$ and $A_-$ are the characteristic inverse length-scale of the $N$ particle bound state on the right and left side of $x=0$ respectively. 
As $a \to 0$ the size of the bound state thus diverges as $\sim 1/a$, consistent with the absence of a stationary state for $a=0$.
\\

We have thus determined the full solution for the density of particles $\rho_s(x)=\partial_x r(x)$, up to two integration constants $C_+$ and $C_-$. We thus need two additional boundary conditions. As we will see below, the choice of these boundary conditions is highly non-trivial and depends on the parameters, leading to different regimes. In particular, the constants $A_\pm$ should be positive so that the density correctly decays at infinity. The condition $A_+>0$ requires that $a<|v_0-b/2|$. If this condition is not satisfied, the density is restricted to the interval $(-\infty,0]$ and a shock appears at $x=0$. Similarly, $A_->0$ requires that $a<v_0+b/2$. For $a>v_0+b/2$, the density will thus be a single delta peak at $x=0$ (phase IV in Fig.~\ref{phase_diagram_nonreciprocal}), which was to be expected since in this case since the noise and the interaction are not strong enough to compete with the confining potential. 

One way to check if a solution is consistent is to consider the total force on a particle at a given position. In particular, to know if a cluster is present at $x=0$ and if the particles can access the regions $x>0$ and $x<0$, one should compute the force at $x=0^-$, $x=0^+$ and exactly at $x=0$. Let us denote $f_\pm(x)$ the total force on a $\pm$ particle at position $x$ and let us write its expression near $x=0$ as a function of the rank fields $h_\pm(x) = \int_{-\infty}^x \rho_\pm(y)dy-\frac{1}{4}$. For a $+$ particle it reads
\bea
&&f_+(0^+) = v_0 - a - b \int_{-\infty}^{0^+} \rho_-(y)dy + b \int_{0^+}^{+\infty} \rho_-(y)dy 
= v_0 - a - 2 b h_-(0^+) \;, \nonumber \\
&&f_+(0) = v_0  - b \int_{-\infty}^{0^-} \rho_-(y)dy + b \int_{0^+}^{+\infty} \rho_-(y)dy = v_0  - b ( h_-(0^+) + h_-(0^-) ) \;, \label{force_pos} \\
&&f_+(0^-) = v_0 + a - b \int_{-\infty}^{0^+} \rho_-(y)dy + b \int_{0^+}^{+\infty} \rho_-(y)dy = v_0 + a - 2 b h_-(0^-) \;, \nonumber
\eea  
and similarly for a $-$ particle,
\bea
&&f_-(0^+) = - v_0 - a + 2 b h_+(0^+) \;, \nonumber \\
&&f_-(0) = - v_0  + b ( h_+(0^+) + h_+(0^-) ) \;, \label{force_neg} \\
&&f_-(0^-) = - v_0 + a + 2 b h_+(0^-) \;. \nonumber
\eea
A cluster of $+$ particles forms at $x=0$ if either $f_+(0)<0$ and $f_+(0^-)>0$, or if $f_+(0)>0$ and $f_+(0^+)<0$ (and similarly for $-$ particles). Another sufficient condition is of course $f_\pm(0^-)>0$ and $f_\pm(0^+)<0$. In addition, the density vanishes on the half line $x>0$ iff ($f_+(0)<0$ or $f_+(0^+)<0$) and ($f_-(0)<0$ or $f_-(0^+)<0$) (and symmetrically for $x<0$). One can see that, when $a>v_0+b/2$, since $|h_\pm(x)|<1/4$, one necessarily has $f_\pm(0^+)<0$ and $f_\pm(0^-)>0$, and thus all the particles remain at $x=0$ (phase IV). Of course, in general the forces $f_\pm(x)$ can only be computed a posteriori from the rank fields, but we will use them to check the consistency of our solutions.
\\

Below we will consider in order the 3 non-trivial phases of Fig.~\ref{phase_diagram_nonreciprocal}: phase I, for $a<v_0-b/2$, where there is no shock, phase II, for $|v_0-b/2|<a<v_0+b/2$, where there is a shock at $x=0$ and the density is restricted to $(-\infty,0]$, and phase III, for $a<b/2-v_0$, where there is a shock at $x=0$ but the density is again supported on the whole real axis.

\subsubsection{Phase I: Absence of shock}

Let us first assume that the  there is no shock, i.e. that $r(x)$ and $s(x)$ are continuous at $x=0$. Using Eq.~\eqref{rel_sr}, this leads to the condition
\be \label{r0s0}
r(0)= \frac{b}{4 v_0} \quad , \quad s(0) = -\frac{a}{2 v_0} \;.
\ee 
The fraction of particles on the negative side is $r(0)+ \frac{1}{2} = \frac{1}{2} + \frac{b}{4 v_0}$. 
Hence for $b>0$ there are more particles on the negative part of the real axis.
Note that \eqref{r0s0} implies that
\be 
h_+(0) = \frac{r(0)+s(0)}{2} = \frac{b-2 a}{8 v_0} \quad , \quad h_-(0) = \frac{r(0)-s(0)}{2} = \frac{b+2 a}{8 v_0} \;,
\ee 
and remember that each field varies in  $[-1/4,1/4]$. In particular, the condition $h_-(0)<1/4$ implies the following condition for the absence of shocks:
\be \label{cond1}
a<v_0-b/2 \;.    
\ee 
Note that since $a>0$, one must also have $b < 2 v_0$. The other bounds on $h_\pm(0)$ imply $a<v_0+b/2$, which is less restrictive for $b>0$, and $a>b/2-v_0$ which also always holds for $a>0$ and $b < 2 v_0$. As we will see below, when the condition \eqref{cond1} is violated, a shock appears at $x=0$.
From the expression of $h_-(0)$, we expect that all the $-$ particles will be on the left in this regime, since $h_-(0)\to 1/4$ as $a\to v_0-b/2$.

As mentioned above, let us also compute the total force near $x=0$. One has, using that $a<v_0-b/2$
\bea
&& f_+(0^+) = v_0-a-\frac{b}{4v_0}(b+2a) >0 \; , \quad f_-(0^+) = -v_0-a+\frac{b}{4v_0}(b-2a) <0 \;, \nn \\
&& f_+(0) = v_0-\frac{b}{4v_0}(b+2a) >0 \;, \quad \quad \quad \ f_-(0) = -v_0+\frac{b}{4v_0}(b-2a) <0 \;, \\
&& f_+(0^-) = v_0+a-\frac{b}{4v_0}(b+2a) >0 \;, \quad f_-(0^-) = -v_0+a+\frac{b}{4v_0}(b-2a) <0 \;. \nn
\eea
Thus, close to $x=0$, the $+$ particles all move towards the right and the $-$ particles all move towards the left, consistent with our hypotheses that there is no shock at $x=0$ and that the particles have access to the whole real axis.
\\

Evaluating \eqref{eqnr_rx_pos} at $x=0$ and using \eqref{r0s0} we find that in the absence of shock at $x=0$,
\be
C_+ = (2 v_0 - b) ( \frac{b}{2 v_0}  + \log(1-\frac{b}{2v_0}) ) \;.
\ee 
Similarly, evaluating \eqref{eqnr_rx_neg} at $x=0$ and using \eqref{r0s0} we obtain,
\be
C_- =  (2 v_0+b) ( \frac{b}{2 v_0}  -  \log(1+\frac{b}{2v_0}) )  \;.
\ee 
Inserting into \eqref{Wpositive0} and \eqref{Wnegative0}, we obtain the full solution in the regime where $0<b<2 v_0$ and $2 a < 2 v_0 - b$,
\bea \label{Wnoshock_pos}
1 - 2 r(x) &=& \frac{2 v_0-b}{b} W\left(\frac{b}{2 v_0} e^{- A_+ x + \frac{b}{2v_0}}\right)  \quad , \ \ \ \quad A_+=\frac{8 a \gamma}{(2 v_0-b)^2- 4 a^2 } \quad , \quad \text{for } x>0 \;, \\
1 + 2 r(x) &=& -\frac{2 v_0+b}{b} W\left(-\frac{b}{2 v_0} e^{A_- x - \frac{b}{2 v_0}}\right) \quad , \quad A_-=\frac{8 a \gamma}{(2 v_0+b)^2- 4 a^2}\quad , \quad \text{for } x<0 \;. \label{Wnoshock_neg}
\eea
By construction one has $W\left(\frac{b}{2 v_0} e^{ \frac{b}{2v_0}}\right) = \frac{b}{2 v_0} $
which yields $r(0) = \frac{b}{4 v_0}$ as required. 
Note that the lower edge $z=-1/e$ of the first branch of the Lambert function $W(z)$ is only reached as $b \to 2 v_0$.

The total density is obtained by taking a derivative with respect to $x$, leading to,
\bea \label{rhos_nr_pos}
&& \rho_s(x) = A_+ \frac{2 v_0-b}{2 b} \left(1 - \frac{1}{1 + W\left(\frac{b}{2 v_0} e^{- A_+ x + \frac{b}{2v_0}}\right) } \right) \quad , \quad \text{for } x>0 \;, \\
&& \rho_s(x) = A_- \frac{2 v_0+b}{2 b} \left(\frac{1}{1 + W\left(- \frac{b}{2 v_0} e^{A_- x - \frac{b}{2v_0}}\right) } -1 \right) \quad , \quad \text{for } x<0 \;, \label{rhos_nr_neg}
\eea 
where we have used that $W'(z)=\frac{W(z)}{z(1+W(z))}$. One finds
\be 
\rho_s(0^+)= \frac{B_+}{2} \frac{2 v_0 - b}{2 v_0 + b } \quad , \quad \rho_s(0^-)= \frac{B_-}{2} \frac{2 v_0 + b}{2 v_0 - b } \;.
\ee 
The ratio
\be 
\frac{ \rho_s(0^-) }{ \rho_s(0^+)} = \frac{(2 v_0 + b)^2}{(2 v_0 - b)^2} \frac{(2 v_0 - b)^2 - 4 a^2}{(2 v_0 + b)^2-4 a^2} 
= \frac{1 - \frac{4 a^2}{(2 v_0 - b)^2}}{1 - \frac{4 a^2}{(2 v_0 + b)^2}}
\ee 
is smaller than unity for $b>0$. On the other hand the decay is slower on the negative side for $b>0$, since then $A_-<A_+$. As we have already noted above, the fact that $r(0)>0$ means that this second effect dominates and that there are more particles on the negative side than on the positive side. 

\subsubsection{Phase II: Shock at $x=0$ and no particles for $x>0$}

Considering the solution \eqref{Wnoshock_pos}-\eqref{Wnoshock_neg}, we noticed above that two things happen as $a \to (v_0 - b/2)^-$. First, $A_+$ diverges, suggesting that the density vanishes on the positive side of the real axis. Second, $h_-(0)\to 1/4$, suggesting that all $-$ particles are located at $x<0$. However, $h_+(0)$ does not converge to $1/4$ in that limit (it converges to $\frac{1}{4} - \frac{a}{2 v_0}$), which suggests the existence of a cluster (i.e. a delta peak) of $+$ particles at $x=0$ for $a>v_0-b/2$ (of weight $a/(2v_0)$ on the frontier between phase I and II). 

Let us thus look for a solution in the regime $a>v_0-b/2$, assuming that the density vanishes for $x>0$ (thus $h_+(0^+)=h_-(0^+)=1/4$). In this case we only need to determine the constant $C_-$ and the weight of the delta peak at $x=0$. In addition, let us assume that $h_-(0^-)=1/4$ remains frozen, meaning that there are no $-$ particles in the shock at $x=0$. Using Eq.~\eqref{rel_sr}, which remains valid as long as there is no shock outside of $x=0$, we obtain
\be 
h_+(0^-)= \frac{1}{4} - \frac{2 a}{b + 2 v_0+2a} \;,
\ee 
and thus,
\be \label{r0s0_shock}
r(0^-)= \frac{b + 2 v_0 - 2 a}{2 (b + 2 v_0 + 2 a)} = \frac{1}{2} - \frac{2 a}{b + 2 v_0+2a} \quad , \quad s(0^-)=- \frac{2 a}{b + 2 v_0 + 2 a} \;.
\ee 
Note that $h_+(0^-)\to-1/4$ as $a\to v_0+b/2$, consistent with the fact that all particles remain at $x=0$ for $a>v_0+b/2$, i.e. in Phase IV. 
Note also that $h_+(0^-)\to 1/4-a/(2v_0)$ as $a\to v_0-b/2$ (for $b<2v_0$), meaning that the weight of the delta peak does not vanish continuously as one approaches this line (i.e. the boundary between the phases I and II). 

Once again, let us compute the total force near $x=0$. One has,
\bea
&& f_+(0^+) = v_0-a-\frac{b}{2} \; , \quad f_-(0^+) = -v_0-a+\frac{b}{2} \; , \nn \\
&& f_+(0) = v_0-\frac{b}{2} \;, \quad \quad \quad \ f_-(0) = -v_0+\frac{b}{2} - \frac{2ab}{b+2v_0+2a} \;, \\
&&f_+(0^-) = v_0+a-\frac{b}{2} \;, \quad f_-(0^-) = -v_0+a+\frac{b}{2} - \frac{4ab}{b+2v_0+2a} \;. \nn
\eea
Let us study the sign of these forces in detail. First, as long as the condition $a>|v_0-b/2|$ is satisfied, one has $f_+(0^+)<0$ and $f_+(0^-)>0$, implying that a cluster of $+$ particles is indeed present at $x=0$. In addition, in this region one has $f_-(0^+)<0$. Thus, no particle can access the region $x>0$, in agreement with our assumptions. In fact, as long as $-v_0+b/2<a<v_0+b/2$, $f_-(x)$ is strictly negative in the whole vicinity of $x=0$, which means that for $|v_0-b/2|<a<v_0+b/2$ all the $-$ particles remain in the region $x<0$ (and do not contribute to the cluster at $x=0$), as we have assumed. Of course, since the $-$ particles can turn into $+$ particles, this implies that some $+$ particles will also be present in this region. However, once they have reached $x=0$ they remain their until they switch sign again.

The sign of the forces $f_\pm(x)$ in the vicinity of $x=0$ is thus compatible with the assumptions of this section (no particles for $x>0$ and no $-$ particles at $x=0$) in the region $|v_0-b/2|<a<v_0+b/2$, which defines phase II and which is also the region where $A_->0$ and $A_+<0$. 
Let us discuss the limits to the phases IV, I and III respectively.
For $a>v_0+b/2$, i.e. phase IV, one has $f_-(0^-)<0$ and all the particles remain at $x=0$. For $b<2v_0$ and $a<v_0-b/2$, i.e. phase I, one finds that $f_+(0^+)>0$ and $f_+(0)>0$. Thus the $+$ particles do not remain confined at $x=0$ and can access the region $x>0$. This means that the correct solution in this region is indeed the one obtained in the previous section, in the absence of shocks. Finally, and perhaps more surprisingly, for $b>2v_0$ and $a<-v_0+b/2$, i.e. in phase III, one finds that $f_+(x)$ is strictly negative and $f_-(x)$ is strictly positive in the whole vicinity of $x=0$, implying that the particles have again access to the whole real axis. In this case, the present assumptions are thus incorrect. This last case will be studied in the next section.
\\

We now determine the full solution for the density in the region $|v_0-b/2|<a<v_0+b/2$, where the present assumptions are valid.  Eq.~\eqref{eqnr_rx_neg} remains true even with a shock at $x=0$, but now the constant $C_-$ is determined by \eqref{r0s0_shock}. This gives
\be 
C_- = \left(b+2 v_0\right) \left(\frac{2
   b}{2 a+b+2 v_0}-\log
   \left(\frac{2 \left(b+2
   v_0\right)}{2 a+b+2
   v_0}\right)\right) \;.
\ee 
This leads to, for $x<0$, using \eqref{Wnegative0} 
\be \label{Wnegative}
1 + 2 r(x) =
 -\frac{2 v_0+b}{b} W\left(-\frac{2 b}{2 a + b + 2 v_0} e^{A_- x - \frac{2 b}{2 a + b + 2 v_0}}\right) \quad , \quad A_-=\frac{8 a \gamma}{(2 v_0+b)^2- 4 a^2}
\;.
\ee
Taking the derivative we find the total density in phase II
\be \label{rhos_nr_neg2}
\rho_s(x) = A_- \frac{2 v_0+b}{2 b} \left(\frac{1}{1 + W\left(- \frac{2b}{2a+b+2 v_0} e^{A_- x - \frac{2b}{2a+b+2v_0}}\right) } -1 \right) \quad , \quad x<0 \;,
\ee
and we recall that $\rho_s(x)=0$ for $x>0$. The smooth part of the density reaches 
a finite limit $\rho_s(0^-)=(b+2 v_0) A_-/(2 a - b + 2 v_0)$ at $x=0^-$.
Furthermore, there is a delta peak at $x=0$ (which contains only $+$ particles) with weight
\be \label{r0s0_shock}
\frac{1}{2}-r(0^-) = \frac{2 a}{b + 2 v_0+2a} \;.
\ee 




\subsubsection{Phase III: Shock at $x=0$}

We now turn to the last region, where $b/2>v_0+a$. This corresponds to the regime where the repulsion exerted by the $+$ particles on the $-$ particles is strong enough to allow the $-$ particles to access the region $x>0$. Indeed in this regime, our numerical simulations show the presence of a cluster of $+$ particles at $x=0$ and a non-zero density on the whole real axis. This suggests that the $+$ particles remain stuck at $x=0$ while the $-$ particles located at $x=0$ can go either to the left or to the right depending on the fluctuations. The absence of $-$ particles in the cluster suggests to impose the continuity of $h_-(x)$ at $0$, i.e. $h_-(0^-)=h_-(0^+)=h_-(0)$. The computation of the two integration constants requires an additional boundary condition. This can be obtained by considering the forces near $x=0$. The fact that $-$ particles are able to go either to the right or to the left requires $f_-(0^+)>0$ and $f_-(0^-)<0$, but also that the force exactly at $x=0$ vanishes
\be 
f_-(0)= - v_0 + b ( h_+(0^+) + h_+(0^-)) = 0 \;. \label{f0cond}
\ee 

Rewriting \eqref{rel_sr},in terms of $h_\pm(x)$, we have
\bea 
&& h_+(0^+)- h_-(0^+) = \frac{a}{b-2 v_0} (1- 2 (h_+(0^+)+ h_-(0^+))) \;, \\
&& h_+(0^-)- h_-(0^-) = -  \frac{a}{b+ 2 v_0} (1 + 2 (h_+(0^-)+ h_-(0^-))) \;,
\eea  
which can be rewritten as 
\bea 
&& h_+(0^+) = h_-(0^+) + \frac{a}{2 a + b - 2 v_0} (1 - 4 h_-(0^+)) \;, \\
&& h_+(0^-) = h_-(0^-) - \frac{a}{2 a + b + 2 v_0} (1 + 4 h_-(0^-)) \;.
\eea 
Using the continuity of $h_-(x)$ and the condition \eqref{f0cond}, we obtain 
\bea  \label{h0_largeb}
&& h_-(0)= \frac{v_0}{2 b} \frac{b^2 + 4 a^2 -4 v_0^2}{b^2 - 4 a^2 - 4v_0^2} \;, \nn\\
&& h_+(0^+) = \frac{\frac{4 a^2 (b-2 a)}{4
   a^2-b^2+4 v_0^2}+2 a+v_0}{2b} 
    \;, \\
&& h_+(0^-) = \frac{\frac{4 a^2 (2 a-b)}{4
   a^2-b^2+4 v_0^2}-2 a+v_0}{2
   b} \;. \nn 
\eea  
Note that $h_-(0)>0$ so that there are more $-$ particles (and thus also more $+$ particles) on the left than on the right. One can check that on the line $b/2=v_0+a$ one has $h_-(0)=1/4$ and $h_+(0^+)=1/4$. In addition, one can check that
\bea 
&& f_+(0^+) = v_0 - a - 2 b h_-(0) <0 \quad , \quad \quad \ f_+(0^-) = v_0 + a - 2 b h_-(0) > 0 \;, \\
&& f_-(0^+) = - v_0 - a + 2 b h_+(0^+) >0 \quad , \quad  f_-(0^-) = - v_0 + a + 2 b h_+(0^-) < 0 \;,
\eea 
which validates our assumptions.
\\

Eq.~\eqref{h0_largeb} leads to
\bea
&& 1-2r(0^+) = \frac{\left(b^2-4 v_0^2\right)
   \left(b-2 a-2
   v_0\right)}{b \left(b^2-4a^2-4 v_0^2\right)} \;, \\
&& 1+2r(0^-) =   \frac{\left(b^2-4 v_0^2\right)
   \left(b-2 a+2v_0\right)}{b \left(b^2-4a^2-4 v_0^2\right)} \;,
\eea
which, inserting into \eqref{eqnr_rx_pos} and \eqref{eqnr_rx_neg} for $x=0$ gives
\bea
&& C_+ 
= \frac{\left(b^2-4 v_0^2\right)
   \left(b-2 a-2v_0\right)}{b^2-4 a^2-4v_0^2}+\left(2 v_0-b\right)
   \log
   \left(\frac{\left(b^2-4
   v_0^2\right) \left(b-2 a-2v_0\right)}{b \left(b^2-4a^2-4
   v_0^2\right)}\right) \;, \\
&& C_- 
= \frac{\left(b^2-4 v_0^2\right)
   \left(b-2 a+2
   v_0\right)}{b^2-4 a^2-4v_0^2}-\left(b+2 v_0\right)
   \log
   \left(\frac{\left(b^2-4
   v_0^2\right) \left(b-2 a+2v_0\right)}{b \left(b^2-4a^2-4
   v_0^2\right)}\right) \;.
\eea
This leads to the solution, for $a<-v_0+b/2$,
\bea
&& \hspace{-1.5cm} 1 - 2 r(x) = \frac{2 v_0-b}{b} W\left(-B_+ e^{- A_+ x - B_+}\right)  \;, \quad A_+=\frac{8 a \gamma}{(2 v_0-b)^2- 4 a^2 } \;, \quad B_+ = \frac{(b+2v_0)(b-2a-2v_0)}{b^2-4a^2-4v_0^2} \;, \\ 
&& \hspace{-1.5cm} 1 + 2 r(x) = -\frac{2 v_0+b}{b} W\left(-B_- e^{A_- x - B_-}\right) \;, \quad A_-=\frac{8 a \gamma}{(2 v_0+b)^2- 4 a^2} \;, \quad B_- = \frac{(b-2v_0)(b-2a+2v_0)}{b^2-4a^2-4v_0^2} \;,
\eea 
Finally, the total density in phase III reads
\bea \label{rhos_shock_pos}
&& \rho_s(x) = A_+ \frac{b-2 v_0}{2 b} \left(\frac{1}{1 + W\left(-B_+ e^{- A_+ x - B_+}\right) } - 1 \right) \quad , \quad x>0 \;, \\
&& \rho_s(x) = A_- \frac{2 v_0+b}{2 b} \left(\frac{1}{1 + W\left(- B_- e^{A_- x -B_-}\right) } -1 \right) \quad , \quad \ \; x<0 \;, \label{rhos_shock_neg}
\eea 
with a delta peak at $x=0$ with weight
\be
h_+(0^+) - h_+(0^-) = \frac{2 a \left(b^2-2 a b-4v_0^2\right)}{b \left(b^2-4a^2-4 v_0^2\right)} \;.
\ee

We compared our analytical results to numerical simulations for finite $N$ in Fig.~\ref{plots_nonreciprocal} (see Sec.~\ref{sec:mainresults} for a discussion). Note that in phase III case, the convergence with $N$ is slower than in the other regimes (see Fig.~\ref{plots_nonreciprocal}) due to the particular role played by the fluctuations.
\\

\section{Conclusion}

In this paper, we studied two different models of run-and-tumble particles with 1D Coulomb interaction: the active jellium model, with a harmonic external potential, and the non-reciprocal active self-gravitating gas, with a non-reciprocal rank interaction and a linear confining potential. Both models converge at large time to a steady state which exhibits a number of phase transitions. In both cases, we achieved significant analytical progress in the computation of the stationary particle density in the large $N$ limit, allowing for a thorough understanding of the different phases of the two models.

Concerning the non-reciprocal self-gravitating gas, there is a lot of space left for extensions of the model. First, one could consider different confining potentials or study the large time behavior in the absence of confinement. Second one could try to generalize the present computation by adding a reciprocal part to the interaction on top of the non-reciprocal component. This would allow to test whether the shocks which occur at the edge for purely a reciprocal rank interaction \cite{activeRDshort} are modified in this case. 
Finally, it would be interesting to extend the present model to a case where the non-reciprocal interaction takes place between two distinct species of particles, independently of their velocity state. Of course, we expect the behavior to be very different in this case. More generally, despite their relevance for biological systems, few exact results have been obtained until now for models of active particles with non-reciprocal interactions. We hope that the present paper will provide tools for more analytical studies of such models in the future.
\\

%
%

{\bf Acknowledgments. }
We acknowledge support from ANR Grant No. ANR- 23-CE30-0020-01 EDIPS.

\appendix

\section{"Vision cone" model} \label{app:vision_cone}

Let us consider another non-reciprocal extension of the active rank diffusion model, inspired from models with "vision cones". Each particle now only receives a force only from the particles which are in front of it (i.e. on the right for $+$ particles and on the left for $-$ particles). The equations of motion read,
\be 
\dot x_i = -\sigma_i(t) \frac{2\kappa}{N} \sum_{j\neq i} \Theta(\sigma_i(t)(x_j-x_i)) -V'(x_i) + v_0\sigma_i(t) \;,
\ee
where $\Theta(x)$ is the Heaviside function and we choose the convention $\Theta(0)=1/2$. Writing $\Theta(x)=\frac{1}{2}(1+{\rm sgn}(x))$, this can be rewritten as
(up to a term $\sigma_i \kappa/N$ which we neglect) 
\bea
\dot x_i &=& \sigma_i(t) \big[ v_0-\kappa-\frac{\kappa}{N} \sum_{j\neq i} {\rm sgn}(\sigma_i(t)(x_j-x_i))\big] -V'(x_i) \\
&=& \sigma_i(t) ( v_0-\kappa ) -\frac{\kappa}{N} \sum_{j\neq i} {\rm sgn}(x_j-x_i) -V'(x_i) \;.
\eea
Quite surprisingly, this is exactly the same equation as the reciprocal active rank diffusion model \eqref{langevin1} with a mapping $v_0\to \tilde v_0 = v_0-\kappa$. The non-reciprocity does not play any particular role in this case and the phase diagram can be directly deduced from the one that we have obtained in the reciprocal case, in \cite{activeRDshort} for $V(x)=0$ and $V(x)=a|x|$, and in the present paper for $V(x)=\frac{\mu}{2}x^2$. Note however that here the effective driving velocity $\tilde v_0$ may be negative, but changing the sign of the velocity simply amounts to exchanging $+$ and $-$ particles so that the full phase diagrams can be obtained by symmetry. This exchange of $+$ and $-$ particles only occurs in the repulsive case $\kappa>0$ and is due to the fact that a $-$ particle located far at the right will "see" the other particles and be repulsed towards the right, while a $+$ particle will feel no interaction (and vice-versa at the left edge). More precisely, the phase diagrams remain the same up to a replacement of $v_0$ by $|\tilde v_0|$ (we only need to recall that $+$ and $-$ particles are exchanged whenever $\tilde v_0<0$). Some regions of the phase diagram will however become inaccessible. In the repulsive case, one has $\tilde v_0/\kappa = v_0/\kappa-1 \in [-1,\infty)$ (or $\kappa/\tilde v_0 = \frac{1}{v_0/\kappa-1} \in (-\infty,-1)\cup(0,\infty)$), thus the entire diagram is still accessible. But in the attractive case, $\tilde v_0/\bar \kappa = v_0/ \bar \kappa+1 \in [1,\infty)$ (or $\bar \kappa/\tilde v_0 = \frac{1}{v_0/ \bar \kappa+1} \in (0,1)$), which means that some regions  of the diagram do not exist in this model. In particular, in the absence of external potential, in the attractive case 
it implies that the density is always smooth, with unbounded support (i.e. the phase $I_s$ in \cite{activeRDshort}) and there is no transition (in the repulsive case, the solution is again an expanding plateau, with support $[-\kappa t,\kappa t]$). In the case of a linear or harmonic confinement however, all the different phases continue to exist, although for some of them in a reduced domain.



\end{document}